\documentclass[apj]{emulateapj}
\usepackage{mathptmx}

\newcommand  \kms      {\ifmmode {\rm km\,s}^{-1} \else km\,s$^{-1}$\fi}
\newcommand  \cc       {\hbox{cm$^{-3}$}}
\newcommand  \cmii     {\hbox{cm$^{-2}$}}
\newcommand  \ergs     {\ifmmode {\rm erg\,s}^{-1} \else erg s$^{-1}$\fi}
\newcommand  \ergcms   {\ifmmode {\rm erg\,cm}^{-2}\,{\rm s}^{-1}
                        \else erg\,cm$^{-2}$\,s$^{-1}$\fi}
\newcommand  \ergcmsA {\ifmmode{\rm erg\,cm}^{-2}\,{\rm s}^{-1}\,{\rm\AA}^{-1}
                        \else erg\,cm$^{-2}$\,s$^{-1}$\,\AA$^{-1}$\fi}
\newcommand \ergcmsHz {\ifmmode{\rm erg\,cm}^{-2}\,{\rm s}^{-1}\,{\rm Hz}^{-1}
                        \else erg\,cm$^{-2}$\,s$^{-1}$\,Hz$^{-1}$\fi}
\newcommand  \phcms    {\ifmmode {\rm ph\,cm}^{-2}\,{\rm s}^{-1}
                        \else ,ph\,cm$^{-2}$\,s$^{-1}$\fi}
\newcommand  \phcmsA   {\ifmmode {\rm ph\,cm}^{-2}\,{\rm s}^{-1}\,{\rm\AA}^{-1}
                        \else ph\,cm$^{-2}$\,s$^{-1}$\,\AA$^{-1}$\fi}
\newcommand  \mbh      {\ifmmode M_{\rm BH} \else $M_{\rm BH}$\fi}

\def\micron{\ifmmode \mu{\rm m} \else $\mu$m\fi}
\def\kms{\ifmmode {\rm km\,s}^{-1} \else km\,s$^{-1}$\fi}
\def\Hubble{\ifmmode {\rm km\,s}^{-1}\,{\rm Mpc}^{-1}
        \else km\,s$^{-1}$\,Mpc$^{-1}$\fi}
\def\ergsec{\ifmmode {\rm ergs\;s}^{-1} \else ergs s$^{-1}$\fi}
\def\ergscm{\ifmmode {\rm ergs\,s}^{-1}\,{\rm cm}^{-2}
          \else ergs\,s$^{-1}$\,cm$^{-2}$\fi}
\def\ergscmA{\ifmmode {\rm ergs\,s}^{-1}\,{\rm cm}^{-2}\,{\rm \AA}^{-1}
          \else ergs\,s$^{-1}$\,cm$^{-2}$\,\AA$^{-1}$\fi}
\def\ergscmHz{\ifmmode {\rm ergs\,s}^{-1}\,{\rm cm}^{-2}\,{\rm Hz}^{-1}
          \else ergs\,s$^{-1}$\,cm$^{-2}$\,Hz$^{-1}$\fi}
\def\Msun{\ifmmode M_{\odot} \else $M_{\odot}$\fi}
\def\Lsun{\ifmmode L_{\odot} \else $L_{\odot}$\fi}
\def\qo{\ifmmode q_{0} \else $q_{0}$\fi}
\def\Ho{\ifmmode H_{0} \else $H_{0}$\fi}
\def\ho{\ifmmode h_{0} \else $h_{0}$\fi}
\def\qo{\ifmmode q_{0} \else $q_{0}$\fi}
\def\ao{\ifmmode a_{0} \else $a_{0}$\fi}
\def\to{\ifmmode t_{0} \else $t_{0}$\fi}
\def\Halpha{\ifmmode {\rm H}\alpha \else H$\alpha$\fi}
\def\Hbeta{\ifmmode {\rm H}\beta \else H$\beta$\fi}
\def\hb{\ifmmode {\rm H}\beta \else H$\beta$\fi}
\def\Hgamma{\ifmmode {\rm H}\gamma \else H$\gamma$\fi}
\def\Hdelta{\ifmmode {\rm H}\delta \else H$\delta$\fi}
\def\Lya{\ifmmode {\rm Ly}\alpha \else Ly$\alpha$\fi}
\def\Lyb{\ifmmode {\rm Ly}\beta \else Ly$\beta$\fi}
\def\hi{\ifmmode \mbox{{\rm H}\,{\sc i}} \else H\,{\sc i}\fi}

\def\ciii{\ifmmode {\rm C}\,{\sc iii} \else C\,{\sc iii}\fi}

\def\oiii{[O\,{\sc iii}]\,$\lambda5007$}

\newcommand{\Oxs}{[\mbox{O\,{\sc iii}}]~$\lambda$5007}
\newcommand{\Oxf}{[\mbox{O\,{\sc iii}}]~$\lambda$4363}

\def\mgii{\ifmmode {\rm Mg}{\textsc{ii}} \else Mg\,{\sc ii}\fi}

\def\o5007{[O\,{\sc iii}]\,$\lambda5007$}

\def\ne212m {[Ne\,{\sc ii}]\,$12.8 \mu m$}

\def \Lop{$L_{5100}$}
\def \Lbol{$L_{\rm bol}$}
\def \Ledd{$L/L_{\rm Edd}$}

\def  \RNLR        {\hbox{$ {R_{\rm NLR}} $}}

\def  \kms         {\hbox{km s$^{-1}$}}          
\def  \cc          {\hbox{cm$^{-3}$}}

\def  \cmii        {\hbox{cm$^{-2}$}}
\def  \mic         {$\mu$m}
\def  \La          {\ifmmode {\rm Ly}\alpha \else Ly$\alpha$\fi}
\def  \Ka          {\ifmmode {\rm K}\alpha \else K$\alpha$\fi}
\def  \Lb          {\ifmmode {\rm L}\beta \else L$\beta$\fi}
\def  \Ha          {\ifmmode {\rm H}\alpha \else H$\alpha$\fi}
\def  \Hb          {\ifmmode {\rm H}\beta \else H$\beta$\fi}
\def  \Pa          {\ifmmode {\rm P}\alpha \else P$\alpha$\fi}
\def  \CIIIb       {\ifmmode {\rm C}\,{\sc iii]}\,\lambda1909
                     \else C\,{\sc iii]}\,$\lambda1909$\fi}
\def  \CIV         {\ifmmode {\rm C}\,{\sc iv}\,\lambda1549
                     \else C\,{\sc iv}\,$\lambda1549$\fi}
\def  \MgII         {\ifmmode {\rm Mg}\,{\sc ii}\,\lambda2798
                     \else Mg\,{\sc ii}\,$\lambda2798$\fi}
\def  \OVI         {\ifmmode {\rm O}\,{\sc vi}\,\lambda1035
                     \else O\,{\sc vi}\,$\lambda1035$\fi}

\def \spitzer      {{\it Spitzer }}

\def\tv{\ifmmode \tau_V \else $\tau_V$}
\def\Chisq{\ifmmode \chi^{2} \else $\chi^{2}$}

\shorttitle{AGN Dusty Structure}
\shortauthors{Mor et al.}
\slugcomment{Submitted to ApJ}

\begin{document}

\title{Dusty Structure Around Type-I Active Galactic Nuclei:\\ Clumpy Torus Narrow Line Region and Near-Nucleus Hot Dust}

\author{
Rivay Mor,\altaffilmark{1}
Hagai Netzer,\altaffilmark{1}
and Moshe Elitzur\altaffilmark{2}
}

\altaffiltext{1}
{School of Physics and Astronomy and the Wise Observatory, 
The Raymond and Beverly Sackler Faculty of Exact Sciences, 
Tel-Aviv University, Tel-Aviv 69978, Israel}
\altaffiltext{2}
{Department of Physics and Astronomy, University of Kentucky,
Lexington, KY 40506-0055}


\begin{abstract}
We fitted \textit{Spitzer}/IRS $\sim2-35\mu m$ spectra of 26 luminous QSOs in
attempt to define the main emission components.
Our model has three major components: a clumpy torus,
dusty narrow line region (NLR) clouds and a blackbody-like dust.
The models utilize the clumpy torus of Nenkova et al. (2008) and are the first to allow
its consistent check in type-I AGNs.
Single torus models and combined torus-NLR models fail to fit the spectra
of most sources but three component models adequately fit the spectra of all sources.
We present torus inclination, cloud distribution, covering factor and torus mass for all sources
and compare them with bolometric luminosity, black hole mass and accretion rate.
The torus covering factor and mass are found to be correlated with
the bolometric luminosity of the sources.
We find that a substantial amount of the $\sim2-7 \mu m$ radiation originates
from a hot dust component, which likely situated in the innermost
part of the torus. The luminosity radiated by this component and its covering factor
are comparable to those of the torus.
We quantify the emission by the NLR clouds and estimate their distance from the center.
The distances are $\sim700$ times larger than the
dust sublimation radius and the NLR covering factor is about 0.07.
The total covering factor by all components is in good agreement with the known
AGN type-I:type-II ratio.
\end{abstract}


\section{Introduction}
\label{sec_intro}
An obscuring dusty structure surrounding an accreting massive black hole (BH) is believed
to be a common feature of most active galactic nuclei (AGNs).
Since the obscuration of the central region is anisotropic,
sources with small inclination angles to the line of sight
(face-on) would be classified as type-I AGNs and those with large
inclination angles (edge-on) would be classified as type-II AGNs.
In this picture, the bulk of the radiation from the central engine is absorbed by the obscuring
structure and re-emitted mainly in mid-infrared (MIR) wavelengths.
The central obscuration is not necessarily a single component structure
and the exact nature of its different components remains an open question and
provides the motivation to the present work.

A main component of the obscuring structure is believed to be a dusty torus.
The MIR spectral energy distribution (SED) of such torus depends on its dimensions and geometry,
the density distribution and the dust grain properties.
Initial attempts to model such tori assumed smooth density distributions
(e.g. Pier \& Krolik 1992; 1993; Granato \& Danese 1994; Efstathiou \& Rowan-Robinson 1995; van Bemmel
\& Dullemond 2003; Schartmann et al. 2005).
This can explain part of the SED but falls short of providing realistic MIR spectra.
For example, Silicate emission has been detected in Type-2 AGNs (e.g. Sturm et al. 2006; Teplitz et al. 2006)
despite of the the fact that absorption features are expected in edge-on tori.
Several studies suggested that the dusty medium should be clumpy
(e.g. Krolik \& Begelman 1988; Rowan-Robinson 1995; Nenkova et al. 2002; Tristram et al. 2007; Thompson et al. 2009).
The recent works of Nenkova et al. (2002) and Nenkova et al. (2008a; 2008b; hereafter N08)
offer the required formalism to calculate the SED of such tori and a framework to
constrain some of their properties.

Other AGN components can contribute to the observed MIR spectrum of AGNs.
Some of this emission may originate farther from the central radiation source,
at distances exceeding the dimensions of the torus.
Broad-band 10 \mic\ imaging of several nearby AGNs suggests extended MIR continuum source
(Cameron et al. 1993; Bock et al. 2000; Tomono et al. 2001; Radomski et al. 2003; Packham et al. 2005).
Dusty clouds in the narrow line region (NLR) may be the source of such radiation
(Schweitzer et al. 2008; hereafter S08).
Thus, a significant contribution at $\sim$10--30 \mic\ due to components not related to the torus,
must be considered.

\begin{deluxetable*}{lcccc} 
\vspace{-0.5cm}
\tablecolumns{5}
\tablewidth{16cm}
\tabletypesize{\footnotesize}
\tablecaption{The QUEST QSO sample\label{tab_qso_sample}}
\tablehead{\colhead{Object} &\colhead{z} &\colhead{${\rm D_L}$} &\colhead{$\log {\rm L_{5100}}$} &\colhead{NIR Data}\\
\colhead{} &\colhead{} &\colhead{(Mpc)} &\colhead{(\ergs)} &\colhead{References}}
\startdata
PG 2349-014 & 0.1740 & 840  & 44.81  & 1,2,3,4,5,6\\
PG 2251+113 & 0.3255 & 1706 & 45.63  & 3,4,7,8\\
PG 2214+139 & 0.0658 & 295  & 44.40  & 2,4,6,7,9,10,11,12\\
PG 1700+518 & 0.2920 & 1504 & 45.68  & 3,4\\
PG 1626+554 & 0.1330 & 623  & 44.44  & 4\\
PG 1617+175 & 0.1124 & 520  & 44.29  & 3,4\\
PG 1613+658 & 0.1290 & 603  & 44.70  & 1,2,3,4,7,11,13\\
PG 1448+273 & 0.0650 & 291  & 44.283 & 2,3\\
PG 1440+356 & 0.0791 & 357  & 44.22  & 3,4,10,13\\
PG 1435-067 & 0.1260 & 589  & 44.39  & 7\\
PG 1426+015 & 0.0865 & 393  & 44.44  & 2,3,4,7\\
PG 1411+442 & 0.0896 & 408  & 44.31  & 3,4,13\\
PG 1309+355 & 0.1840 & 891  & 45.081 & 4\\
PG 1302-102 & 0.2784 & 1424 & 45.17  & 4,14\\
PG 1244+026 & 0.0482 & 213  & 43.593 & 2,7\\
PG 1229+204 & 0.0630 & 281  & 43.895 & 1,2,3,4,13,15,16\\
PG 1126-041 & 0.0600 & 267  & 43.82  & 1,2,4,6,13,16\\
PG 1116+215 & 0.1765 & 851  & 45.13  & 4,6,7,17\\
PG 1004+130 & 0.2400 & 1201 & 45.42  & 3,4,8,14,18\\
PG 1001+054 & 0.1605 & 766  & 44.55  & 4,8,13,14\\
PG 0953+414 & 0.2341 & 1168 & 45.11  & 3,4\\
PG 0838+770 & 0.1310 & 613  & 44.16  & 2,3,4,13\\
PG 0026+129 & 0.1420 & 670  & 44.66  & 3,4,7,8,14,18\\
PG 0157+001 (Mrk 1014) & 0.1630 & 779 & 44.68 & 2,3,4,5,16,19,20,21\\
PG 0050+124 (IZw1)     & 0.0611 & 273 & 44.30 & 6,7,9,10,12,16,22,23,24\\
B2 2201+31A & 0.295  & 1522 & 45.91 & 4,8
\enddata
\tablerefs
{
(1) Rudy et al. (1982);
(2) 2MASS magnitudes (Jarrett et al. 2000);
(3) Neugebauer et al. (1987);
(4) Guyon et al. (2006);
(5) Glikman et al. (2006);
(6) Sanders et al. (1989);
(7) Elvis et al. (1994);
(8) Neugebauer et al. (1979);
(9) Balzano \& Weedman (1981);
(10) Rieke (1978);
(11) McAlary et al. (1983);
(12) Stein \& Weedman (1976);
(13) Surace et al. (2001);
(14) Hayland \& Allen (1982);
(15) Gavazzi \& Boselli (1996);
(16) Veilleux et al. (2006);
(17) Matsuoka et al. (2005);
(18) Sitko et al. (1982);
(19) Scoville et al. (2000);
(20) Surace et al. (1999);
(21) Imanishi et al. (2006);
(22) Spinoglio et al. (1995);
(23) Jarret et al. (2003);
(24) Allen (1976)
}
\end{deluxetable*}

Another component not necessarily related to the torus is hot dust
emission in the immediate vicinity of the central engine.
Minezaki et al. (2004) reported delayed and correlated variations between
the V and the K band emission in the nucleus of the Seyfert 1 galaxy NGC4151.
The measured lag between the V and K bands, $\sim48$ days,
lead to the conclusion that the near infrared emission is dominated
by thermal radiation from hot dust $\sim$0.04 pc from the center.
This result was recently supported by the work of Riffel et al. (2009)
who fitted the near infrared spectrum
of NGC 4151 with $\sim1300 K$ blackbody (BB) spectrum representing
emission from hot dust in the inner region of the torus.
Other studies used similar models to fit the SED of more luminous AGNs (e.g. Edelson \& Malkan 1986;
Barvainis 1987; Kishimoto et al. 2007).
The study of Suganuma et al. (2006) indicates that luminous and variable K-band emission
is common in several nearby Seyfert-1 galaxies.
More generally, observational support for powerful 1--3 \mic\ emission is known for decades
(e.g., Hyland \& Allen 1982; McAlary et al. 1983; Neugebauer et al. 1987).
It is not clear that such emission is consistent with the torus
dust emission, which is expected to peak at longer wavelengths.
Furthermore, part of this radiation requires dust temperature that exceeds
the sublimation temperature of silicate type grains suspected to be responsible
for the bulk of torus MIR radiation.

The high quality spectra made available by the IRS spectrometer on-board
the \textit{Spitzer} Space Observatory (Houck et al. 2004) allows the detailed
analysis of the MIR SED of a large number of AGNs.
Here we fit the observed MIR spectra of 26 PG QSOs, already investigated
by S08, using more realistic three component models
made of  a clumpy torus, dusty NLR clouds and very hot dust clouds.
In \S\ref{sec_sample} we describe the observational data.
In \S\ref{sec_modeling} we detail our model and the fitting procedure
and in \S\ref{sec_Results} we present the results of this procedure.
Our main findings are summarized and discussed in \S\ref{sec_discussion}.

\section{Sample Selection \spitzer\ Observations and Data Reduction}
 \label{sec_sample}
%
Our sample consists of all AGNs in the QUEST \spitzer\ spectroscopy
project (PID 3187, PI Veilleux). It is described in detail in Schweitzer et al.
(2006; hereafter S06), Netzer et al. (2007) and S08.
Most of the objects are Palomar-Green (PG) QSOs (Schmidt \& Green 1983) and are taken from Guyon (2002)
and Guyon et al. (2006). The luminosity range is
\Lop$\approx10^{44.5-46}\,\ergs$ where \Lop\ stands for $\lambda L_{\lambda}$ at rest wavelength 5100\AA.
Radio loudness and infrared excess are typical of these
properties in low redshift PG QSOs. The QUEST sample, while representing the Guyon et al. (2006) sample,
includes fewer sources and thus is not  complete. This was explained in S06. Thus, several of the
correlations discussed below may differ from those in the entire sample in a way which is
difficult to asses until a larger data set, with the same spectral coverage, is
obtained. Here we aims at modeling the entire 2-35 \mic\
SED hence we omit PG 1307+085 (which appeared in the S06 sample) due to
lack of observations in the 5.2-8.7 \mic\ range (\spitzer SL1 mode).
Table \ref{tab_qso_sample} lists names, redshifts, and \Lop\ for all 26 QSOs in our sample.

The \spitzer observations
and the data reduction of the QUEST sample are detailed in S06 and
are summarized here for completion.
IRS spectra for all objects were taken in both low resolution
(5-14 \mic) and high resolution (10-37 \mic) modes.
The standard slit widths of 3".6 to 11".1 include flux from the
QSOs hosts and the vicinity of the AGNs.
Data reduction starts from the basic calibrated data (BCD) provided by the \spitzer pipeline.
Specially developed IDL-based tools are then used for removing outlying values for
individual pixels and for sky subtraction.
The SMART tool (Higdon et al. 2004) is used for extraction of the final spectra.
More details are given in S06.

We have supplemented the \spitzer spectra with near infrared (NIR)
data obtained from the literature. The data are obtained from various sources,
in particular Neugebauer et al. (1987) and the 2MASS extended and point source catalogs (Jarrett et al. 2000).
For some of the sources there are  several photometric measurements.
For these, we average all measurements in each band and
used the standard deviation as the flux error. For single observations,
the flux error is estimated as 20\%. These uncertainties must be underestimated
since many of the NIR data were taken about 20 years before 
the \spitzer observations and all sources are variables.
For a list of JHKL data references see table~\ref{tab_qso_sample}.

We computed the bolometric luminosity of all sources
using their \Lop\ and a simple bolometric correction factor, BC, to convert it to \Lbol.
The best values of BC are discussed in Marconi et al. (2004), Netzer et al. (2007),
and Netzer (2009). The approximation used here is taken from Netzer (2009) and is given by
BC$=9-\log{L_{44}}$, where $L_{44}=$\Lop/$10^{44}$ \ergs. For most objects the value of
\Lop\ is obtained from the observations of Boroson \& Green (1992).
For PG~1244+026, PG~1001+054, and PG~0157+001 we use spectra from the
SDSS (York et al. 2000) data release seven (DR7, Abazajian et al.2008)
that were measured in the way described in Netzer and Trakhtenbrot (2007).

There are two uncertainties associated with the use of \Lbol.
The first is source variability, which is an important
effect since the optical and MIR spectroscopy were separated by many years. We estimate this uncertainty
to be a factor of $\sim1.5$. The second uncertainty involves the approximation used for BC.
We estimate this uncertainty to be $\sim$30\%.
Both uncertainties affect the derived model parameters such as torus covering factor and NLR distance.
Because of this, we do not attach great importance to specific values in specific sources
obtained from our best acceptable models.
On the other hand, the sample is large enough to enable a significant analysis
of its mean properties since the larger of these effects, due to source variability, is changing
in a random way. The uncertainty due to the estimate of BC is smaller but more problematic since
the expression we use can under-estimate or over-estimate \Lbol\, for {\it all sources}. This can
introduce a systematic difference in torus and other component properties, e.g. a smaller covering
factors for all sources. We return
to this point in \S\ref{sec_dis_torus_properties}.


\section{Spectral Modeling}
\label{sec_modeling}
\subsection{Spectral Components}
\label{sec_spectral_comp}

The aim of the present work is to fit the 2-35 \mic\ spectra
of the sources in our sample. For this we combine models of three different physical
components; a dusty torus, a dusty NLR and a very hot dust component.
An additional starburst component representing emission from the host galaxy,
is also present but is not part of the fit procedure. This component is subtracted
from the observed spectra using the MIR spectrum of M82 in a way similar to the one used in S06.
The reality of the subtraction can be judged from the remaining emission or absorption in the wavelengths
corresponding to the strong PAH features (e.g. 6.2 or 7.7 $\mu$m).
More details are given in S06 and in \S\ref{sec_model_fit}.
This section describes the spectral properties of the three components and the following
sections give  detailed account of the fitting procedure.

The first component represents a dusty torus surrounding the central energy source.
AGN unification schemes suggest that the central engine is surrounded by dusty,
optically thick, toroidal structure (e.g., Krolik \& Begelman 1988; Antonucci 1993).
Earlier torus models assumed a smooth density distribution
(Pier \& Krolik 1992; Rowan-Robinson 1995; van-Bemmel \& Dullemond 2003).
They all suffer from various incompletions and their agreement with MIR spectral observations
is generally poor. The more recent clumpy torus models of N08 represent
a significant improvement and are the basis for the present study.

The fundamental difference between clumpy and smooth density distributions
is that the former allows the radiation to propagate freely
between different regions of the optically thick medium.
The clumpy dust distribution results in the coexistence of clouds with
a range of dust temperatures at the same distance from the central radiation source.
They also allow clouds at large distances to be exposed to the direct AGN continuum.
In contrast, smooth density distribution models associate a certain dust
temperature with a certain distance from the central source, thus limiting
the range of acceptable SEDs.
These differences allow the clumpy torus models to have a range of spectral properties
not accessible in smooth density distribution torus models.
N08 describe the formalism and the detailed
radiative transfer used to calculate the MIR spectrum of
clumpy dusty tori and our use of the torus models
follows the same procedure.

The first parameter of the clumpy torus model is the
inner radius of the cloud distribution that is set to the dust
sublimation radius $R_d$. This corresponds to a dust sublimation
temperature $T_{\rm sub}$ that depends on the grain properties and mixture. Here we adopt
two different sublimation radii appropriate for two types of grains. The first is
\begin{equation}
R_{d,C} \simeq 0.5 L_{46}^{1/2} \left (\frac{1800\,\rm{K}}{T_{\rm sub}}\right )^{2.6} \,\rm{pc},
\label{eq:R_d_c}
\end{equation}
where $L_{46}=L_{\rm bol}/10^{46}\,\, {\rm erg s^{-1}}$.
This gives the innermost radius where graphite dust can survive. 
It corresponds to a sublimation temperature of about 1800K and is the one used in S08.
N08 adopted a somewhat larger distance, which is more appropriate
for silicate type grains with a sublimation temperature of 1500K.
This is given by
\begin{equation}
R_{d,Si} \simeq 1.3 L_{46}^{1/2}  \left (\frac{1500\,\rm{K}}{T_{\rm sub}}\right )^{2.6} \, \rm{pc}.
\label{eq:R_d_Si}
\end{equation}
In the following we specify which of the two is used for each purpose.
The clumpy torus model requires six additional parameters:
\begin{enumerate}
\item The visual (5500\AA) dust optical depth of a single cloud, \tv\ (all clouds are assumed to have the same \tv).
\item The mean number of clouds along a radial equatorial line, $N_{0}$.
\item The ratio between the outer dimension of the torus and $R_d$, Y.
\item The inclination angle of the torus with respect to the line of sight, $i$.
\item The torus width parameter $\sigma$, which is analogues to its opening angle.
\item A parameter q that specifies the radial power-law distribution of the clouds, i.e. $N(r) \propto r^{-q}$, where $N$ is the number of clouds.
\end{enumerate}

N08 also define an additional parameter that is set by the above free parameters.
This is the probability $P_{\rm esc}(\beta)$ that light from the central source will escape
the obscuring structure at a given angle $\beta$ without interacting with the clouds.
Assuming that individual clouds are optically thick
\begin{equation}
 P_{\rm esc}(\beta) = e^{-N_{0}e^{-\frac{\beta^{2}}{\sigma^{2}}}},
 \label{eq:Pesc}
 \end{equation}
where $\beta=\pi/2-i$.
By integrating $P_{\rm esc}$ over all angles and subtracting from 1,
we obtain the probability of absorption by the torus. 
\begin{equation}
 f_{2} = 1 - \int_{0}^{\pi/2}P_{\rm esc}(\beta)\cos(\beta)d\beta.
 \label{eq:f2}
 \end{equation}
This is also the fraction of obscured objects in a random sample
(e.g. the fraction of type-II AGNs out of the entire AGN population).
$f_2$ is equivalent to the ``real" (geometrical) covering factor of the torus, and
{\it does not} depend on the inclination angle since it reflects the global torus geometry.
In other words, $f_2$ is the ratio between the total torus luminosity and \Lbol.
Due to the anisotropy of the torus radiation, $f_2$ differs from the
{\it apparent} covering factor of the torus deduced from the ratio between its observed
(angle dependent) luminosity and the bolometric luminosity of the central source.
This apparent covering factor can be represented by
\begin{equation}
 f(i) = \frac{1}{L_{\rm bol}}\int_{2_{\micron}}^{100_{\micron}}L(i,\lambda)d\lambda ,
\label{eq:f(i)}
\end{equation}
where $L(i,\lambda)$ is the angle dependent monochromatic luminosity of the torus.
For a more detailed description of all torus parameters, as well as
other model properties and assumptions, see N08.

The emission and absorption properties of the torus
depend on the dust grains composition and other
properties. All models consider here
assume spherical dust grains with MRN size distribution
(Mathis, Rumple \& Nordsieck 1977). The dust has a standard
galactic mix of 53\% silicates and 47\% graphite. The optical properties of
graphite are taken  from Draine (2003) and for silicates from Ossenkopf, Hennig \& Mathis (1992) (OHM).
The OHM dust mixture produces better agreement with observations of the 10 and 18 \mic\
silicate features (Sirocky et al. 2008) and is the only one used in the present work.

\begin{figure*}[ht] 

 \vspace{1cm}
 \includegraphics[scale=.3]{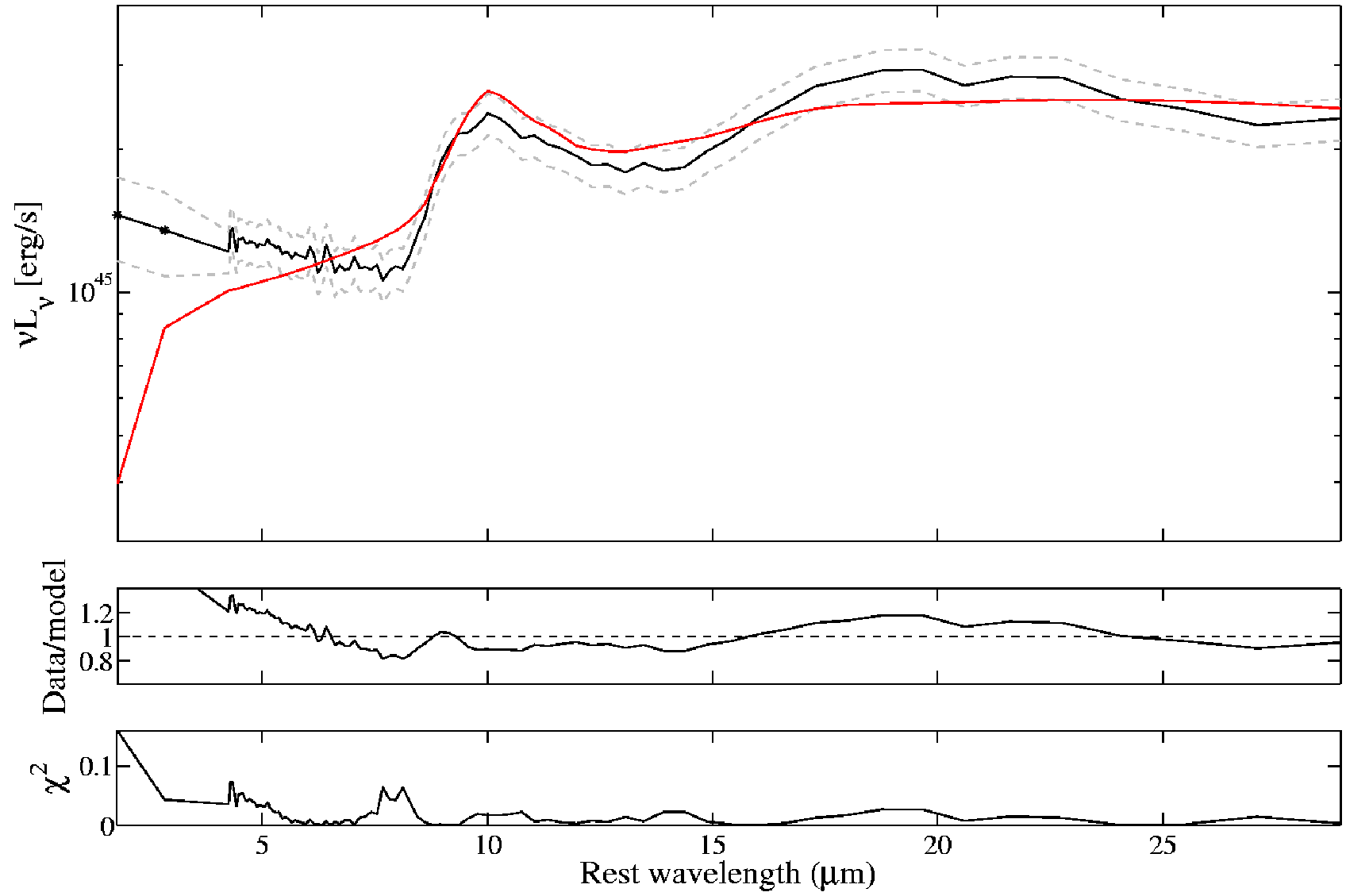}
 \centering
 \includegraphics[scale=.3]{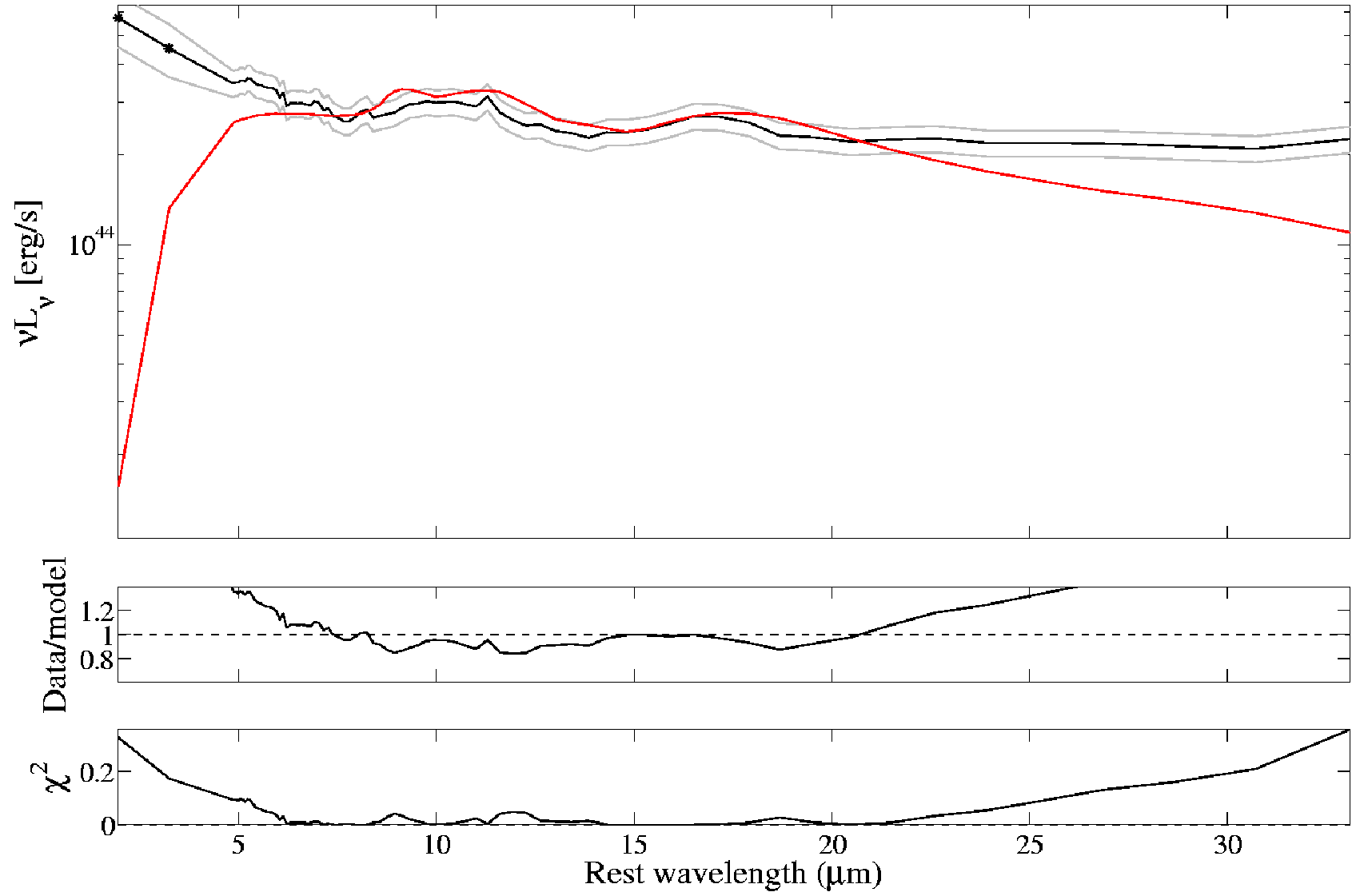}
 \centering
\caption{Best torus-only fit to the spectrum of PG~1004+130 (left; best fit of this type)
and PG~1440+356 (right; typical torus-only fit). The observed binned data are shown in black,
asterisks represent K, L photometry, and the best fit model is shown in red (top panels).
The quality of the fit is demonstrated by the ratio between the data and the model (middle panels)
and the \Chisq value in each wavelength bin (bottom panels).
Note the clear flux deficiencies at short wavelengths in both fits and the
additional flux deficiencies at longer wavelengths for PG~1440+356}
\label{fig:1comp_fit}
\end{figure*}

The second component of the model represents a collection of dusty NLR clouds.
The motivation for this component is explained in S08 where it was shown that such a component
can contribute, significantly, to the MIR flux of luminous AGNs.
The properties assumed here for these clouds are similar to the ones used in S08.
We assume constant column density clouds with $N_{\rm H}=10^{21.5}\,\cmii$.
We further assume constant hydrogen density of 10$^{5}$ \cc, solar composition
and galactic dust-to-gas ratio. The important physical parameters for this component are
the cloud-central source distance
(which determines the dust temperature), the incident SED
and the dust column density. The assumed gas density and column density only serve to define
the emission from this component using convenient parameters such as the ionization parameter.
The NLR component is assumed to be concentrated in a thin spherical shell with a small covering
fraction. These assumptions (that are reviewed later) mean that a \textit{single} thin shell with the chosen
properties is contributing to any 3-component model.

The calculations of the emitted IR continuum of the clouds, as well as
their emission line spectrum, are obtained with the photoionization code
ION (Netzer 2006 and references therein).
The assumed SED of the central continuum is the one described in S08
and the resulting NLR spectra are similar to those
shown in S08 Fig.~1. The clouds are optically thin to MIR radiation and hence
no radiative transfer is required for the calculations of this emission.
Obviously, this is not the case for the UV part of the spectrum where the opacity is much
larger and where the transfer approximation we use is the one built into ION.

The third component is a single BB representing hot dust emission.
The need for such a component has been noticed in earlier works, e.g. in S08.
This is very clear from the data, where a rise
towards short wavelengths can be seen from  the IRS spectra even
without the supplemented NIR photometry.
The simplified BB assumption is not entirely consistent with  hot dust clouds illuminated from one
side since for large dust optical depth, there must be a temperature gradient across such clouds. For
small dust optical depth, the dust temperature is more uniform but the cloud may be partly transparent to the
incident radiation.
For the purpose of the present study, we assume that all the incident
optical-UV radiation is absorbed by this component and the dust temperature is constant. We discuss this component further
in \S\ref{sec_res_3_components} and \S\ref{sec_dis_BB_properties} below.

\subsection{Fitting Method}
\label{sec_model_fit}
As shown in S08 and explained in detail in Netzer et al. (2007),
the starburst contribution to the MIR spectrum can be significant,
especially at long wavelengths.
To remove this contribution, we follow the procedure of S08
who fitted and subtracted a ``nominal'' M82 spectrum from all spectra prior to the model fitting.
We use the ISO-SWS mid-IR spectrum of M82 from Sturm et al. (2000) and
the scaling factors of S08 that were obtained from the intensity of the
strong PAH emission features that are clearly seen in some of our objects.
Although this component is subtracted prior to the model fitting, its
normalization introduces another degree of freedom to the procedure.
The remaining spectra are assumed to represent the intrinsic AGN continuum.

The fitting procedure minimizes the modified \Chisq,
\begin{equation}
\chi^{2} = \frac{1}{N_{\rm dof}}\sum_{\lambda} \left (\frac{\nu L_{\rm \nu,obs}(\lambda)-\nu L_{\rm \nu,model}(\lambda)}{\sigma_{\lambda}}\right)^{2},
\label{eq:chi2}
\end{equation}
where $N_{\rm dof}$ is the number of degrees of freedom, $\nu L_{\rm \nu,obs}(\lambda)$ is the
starburst subtracted luminosity deduced from the observed monochromatic flux density
at rest wavelength $\lambda$, and $\nu L_{\rm \nu,model}(\lambda)$
is the combined emission due to all three components,
\begin{equation}
L_{\rm \nu,model}(\lambda) = L_{\rm \nu,torus}(\lambda)+a_{\rm NLR}L_{\rm \nu,NLR}(\lambda)+a_{\rm BB}L_{\rm \nu,BB}(\lambda),
\label{eq:Fmodel}
\end{equation}
where $a_{\rm NLR}$ and $a_{\rm BB}$ are fitting parameters.
The bolometric luminosity of the source uniquely determine the total emergent flux
of the torus component through the radiative transfer calculation.
Since an independent measure of the bolometric luminosity is available for these type-I sources,
the normalization of the torus component is \textit{not} a free parameter of the fitting procedure (i.e. the first
term in Eq.~\ref{eq:Fmodel} is set by the torus parameters but the total energy emitted by the torus is fixed.

To reduce the uncertainty, we first bin the observed spectra into $\sim$100 bins
of equal energy widths and take in each the mean observed flux.
The torus, NLR and BB models are then interpolated to the same energy
grid. The binning of the data practically removes all emission lines and
smooth the spectrum, allowing to focus on the broad band continuum shape.
Some remaining peaks may result in larger local \Chisq\ values but since none
of the models contains lines, the effect would be the same for all fits.

According to S08, most of the errors in the IRS spectra are systematic
and are approximately ~10\% of the flux.
We add this systematic error to the standard deviation of the data points in each energy bin
to get the value of $\sigma_{\lambda}$ in Eq. \ref{eq:chi2}. This is taken to represent
the error on the observed binned flux.
The use of this $\sigma$ affects the derived \Chisq\, such that
its absolute value is different from the one normally used in statistical analysis.
However, the relative value of \Chisq\ can still be used to discriminate
between different models and will be treated
as such in the remaining of the paper.

The fitting algorithm computes \Chisq\ values for all possible combinations of torus,
NLR and BB models. There are six free parameters in the torus model that describe its
geometrical properties (see section~\ref{sec_spectral_comp}).
The radial extent of the torus, Y, ranges from 5 to 60 $R_d$.
The total number of clouds along radial equatorial lines ranges from 1 to 10.
The optical depth of individual cloud, $\tau_V$, ranges from 5 to 100.
The torus width parameter, $\sigma$, ranges from $15^{\circ}$ to $75^{\circ}$.
The index of the power law radial distribution of clouds, $q$, only accepts one of the three
integer values,  0, 1 or 2.

Finally, the torus inclination angle, $i$, ranges from $0^{\circ}$ to $90^{\circ}$.
In the NLR model there are two free parameters, the cloud distance
and the  normalization factor $a_{\rm NLR}$.
The distance is changed in steps of 0.075 dex between 1 pc to 850 pc
for a source with $L_{\rm bol}=10^{45}\ergs$.
For the hot BB, there are also two free parameters,
temperature and a normalization factor $a_{\rm BB}$.
The range in temperatures is 800--1800~K.

\section{Results}
\label{sec_Results}

\subsection{Torus only models}
\label{sec_res_torus_only}

We first examine the possibility that single torus models can explain the observed
spectrum. This is important in order to evaluate the general
properties of clumpy tori models and has never been attempted
on large  \spitzer-IRS samples. In general, allowing for a torus-only model results
in large \Chisq\ and poor fit to almost all spectra.
On average, \Chisq\ values of such fits are about an order of magnitude
larger than those obtained by using all three components. None of
the torus models could fit well the full observed spectrum of any of the objects.
The best torus-only models are those fitted to the spectra of
PG~1626+554 ,PG~1004+130 and PG~0838+770. These objects are fitted reasonably
well around the silicate emission features and at long wavelengths but fail
at $\lambda \leq 8 \mu$m where they show clear flux deficiencies
(see left panel of Fig.~\ref{fig:1comp_fit}).
Single torus fits to all other objects are significantly worse.
This is illustrated in the right panel of Fig.~\ref{fig:1comp_fit} that shows
the best torus-only fit to the spectrum of PG~1440+356.
Not only that the model poorly fits the silicate features
region, it all shows insufficient flux both at shorter and longer wavelengths.

\subsection{Combined torus-NLR models}
\label{sec_res_torus_and_NLR}

\begin{figure}[ht] 
\vspace{0.3cm}
\epsscale{1.0}
\plotone{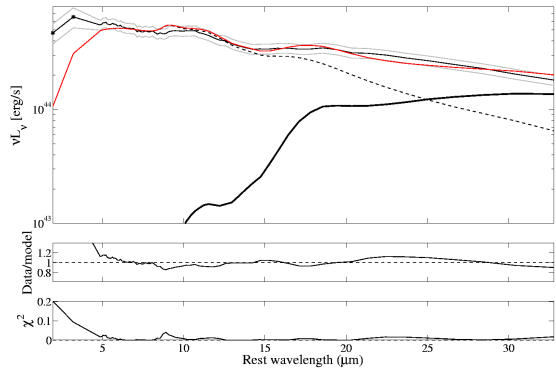}
\caption{Best fit to the spectrum of PG~1411+442 (red line) using a combination
of torus (dashed line) and NLR (thick black line) components. Middle and bottom
panels are as in Fig.\ref{fig:1comp_fit}.}
\label{fig:2comp_fit}
\end{figure}

Next we examine the quality of two component models.
In general, the fits improved compared with torus-only models
especially over the wavelength range $\lambda\gtrsim$6 \mic.
An example is shown in Fig.~\ref{fig:2comp_fit} where we reproduce the best two component fit
to the spectrum of PG~1411+442. In this case, the fit is within 10\% of the
observed flux for all $\lambda\gtrsim$5 \mic. Clear deviations are still found at the short
wavelength part of the spectrum.
We were able to obtain satisfactory fits over the $\sim$6--35 $\mu$m range for
17 out of the 26 objects using two component models.
The \Chisq\ values for these fits are, on average, a factor 6 smaller than the
single torus fit values. These results are comparable to
those shown in S08 where a combination of several BBs and NLR components was
fitted to the 5--35 $\mu$m spectrum of all sources.
Only in PG~1626+554, which is not part of the above group of 17 objects,
the torus-only fit is superior to the two-component fit (i.e. the NLR component
is not required). For the remaining 8 objects, the \Chisq\ values of the two-component
fits are only about a factor 2 smaller than the values obtained for torus-only fits.
In these 8 fits, none of the main spectral features are well fitted.

In all cases fitted with by two component models, a significant fraction of the flux shorter of
$\sim$6 \mic\ is still unaccounted for.
The mean missing fractional (i.e. the difference in flux between model and observations over the 2--35 $\mu$m range) is about 40\%.

\begin{deluxetable}{lccc} 
\tablecolumns{4}
\tablewidth{0pt}
\tabletypesize{\footnotesize}
\tablecaption{\Chisq\, fit results\label{tab_chi2_values}}
\tablehead{\colhead{Object} &\colhead{Torus-only fit} &\colhead{2-component fit} &\colhead{3-component fit}}
\startdata
PG 2349-014 & 6.80 & 2.34 & 0.14 \\ 
PG 2251+113 & 8.99 & 3.95 & 0.25 \\ 
PG 2214+139 & 2.63 & 1.99 & 0.27 \\ 
PG 1700+518 & 2.03 & 0.77 & 0.24 \\ 
PG 1626+554 & 1.10 & 1.12 & 0.50 \\ 
PG 1617+175 & 3.24 & 1.14 & 0.10 \\ 
PG 1613+658 & 2.62 & 1.00 & 0.12 \\ 
PG 1448+273 & 1.56 & 1.24 & 0.31 \\ 
PG 1440+356 & 3.36 & 1.34 & 0.15 \\ 
PG 1435-067 & 3.59 & 0.52 & 0.31 \\ 
PG 1426+015 & 2.15 & 0.93 & 0.11 \\ 
PG 1411+442 & 2.14 & 0.93 & 0.10 \\ 
PG 1309+355 & 3.21 & 1.52 & 0.88 \\ 
PG 1302-102 & 3.08 & 0.70 & 0.10 \\ 
PG 1244+026 & 1.03 & 0.49 & 0.34 \\ 
PG 1229+204 & 2.45 & 0.94 & 0.18 \\ 
PG 1126-041 & 4.86 & 1.51 & 0.24 \\ 
PG 1116+215 & 5.87 & 2.74 & 0.24 \\ 
PG 1004+130 & 1.90 & 0.58 & 0.23 \\ 
PG 1001+054 & 7.91 & 2.23 & 0.28 \\ 
PG 0953+414 & 7.01 & 4.15 & 0.45 \\ 
PG 0838+770 & 1.61 & 0.79 & 0.26 \\ 
PG 0026+129 & 7.62 & 2.07 & 0.31 \\ 
PG 0157+001 & 12.40 & 1.80 & 0.81 \\ 
PG 0050+124 & 8.26 & 1.39 & 0.20 \\ 
B2 2201+31A & 4.61 & 1.19 & 0.09 \\ 
\enddata
\end{deluxetable}

\subsection{Three component models}
\label{sec_res_3_components}

The missing flux in the short wavelength part of the spectrum in most of
our sources is the main motivation for our
third, hot dust component.
This was not discussed in S08 who fitted only the $\lambda \geq 5\mu$m part
of the spectrum.

We have investigated two possibilities: The first is emission due to
hot dust with the same grain composition used for the other components.
This can represent additional dusty clouds in the innermost part of
the torus that are not part of the clumpy structure of N08.
Including such a component, obtained from the single cloud models of N08,
resulted in poorer agreement between model and observations.
The reason is that such a component produces
strong silicate emission features, especially the broad
10 \mic\ feature, that completely change the SED.
The apparent strength of the 10 \mic\ and the 18 \mic\ features in pure emission models represents
the average temperature of the dust. The introduction of very hot silicate dust gives
the impression of a very strong 10 \mic\ feature compared with a weak 18 \mic\ feature,
which clearly contradicts the observations.
Since the hot dust luminosity is very significant, the variations in the
intensities of the other components cannot compensate for this increase.
Thus, using hot silicate dust we could not successfully fit the observations.

The second possibility is a hot BB.
The inclusion of the this component improves, significantly,
the quality of most fits. It allows for a better fit at both
short and long wavelengths since the torus and NLR components
are no longer needed to account for the short wavelength part of the spectrum.
For the same reason, the models come closer to the observations at 10 \mic\ and 18 \mic.
The \Chisq\ values obtained with the three-component fits are, on average,
an order of magnitude smaller than the torus-only fits and about a factor of 3
smaller compared with the two-component fits. \Chisq\ values for all objects and all
types of fits are listed in table~\ref{tab_chi2_values} (note again that these values
do not represent the formal \Chisq\ values).
Given all three components, we obtain fits that are within
20\% of the observations over the entire wavelength range. In the wavelength range
of the silicate emission features ($\sim8-20$ \mic), 19 of our best fit models
deviates by no more than 5\% from the data.
In the remaining 7 sources, the deviation over this wavelength range is at most 10\%.
A more realistic model for the short wavelength component is a
collection of hot graphite-only clouds (see \S\ref{sec_dis_BB_properties}).
Such clouds have no silicate features
in their spectra. Model fitting including such a component is deferred to a future publication.

\begin{figure*}[ht] 
\centering
\includegraphics[scale=.3]{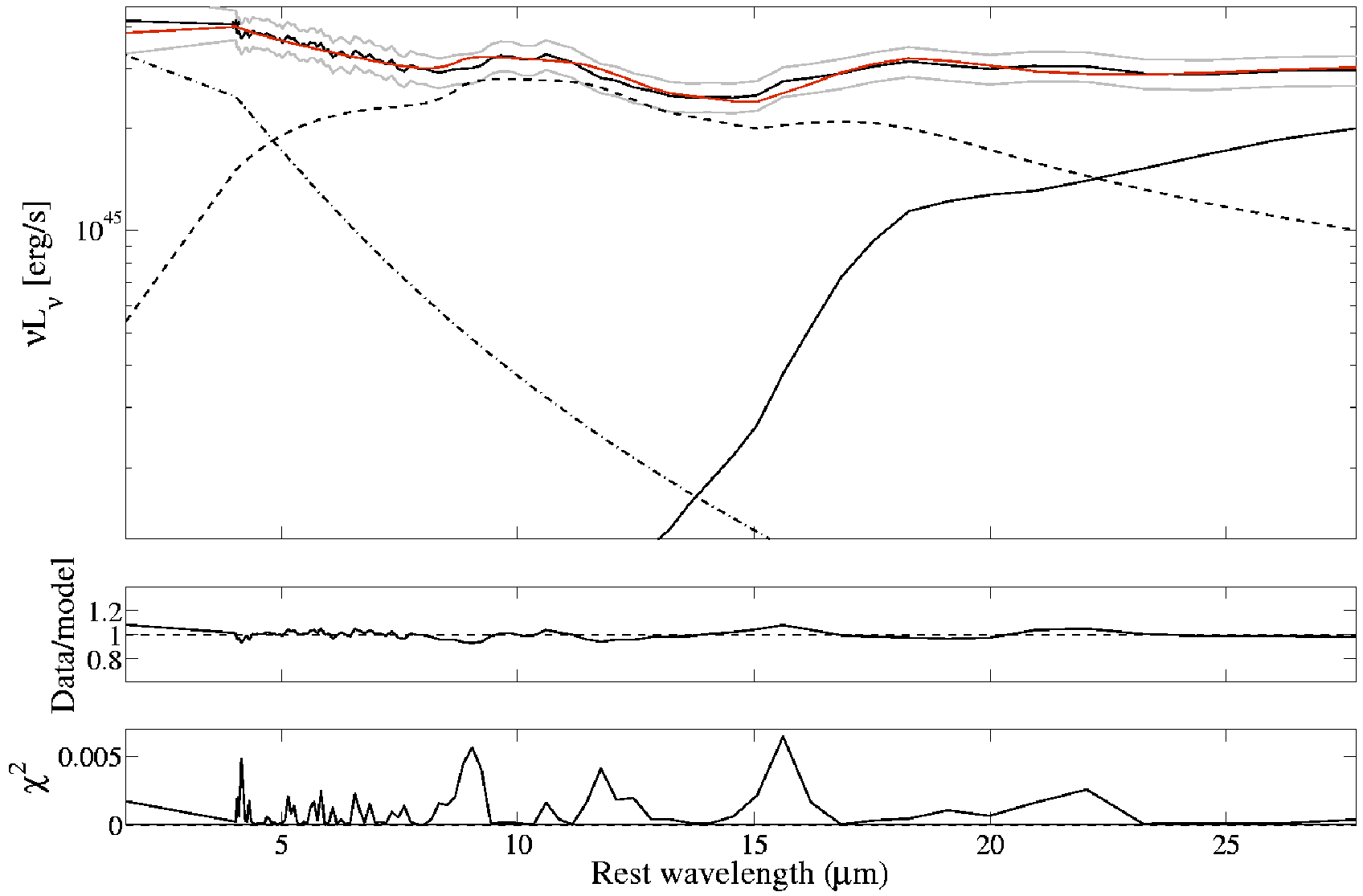}
\centering
\includegraphics[scale=.3]{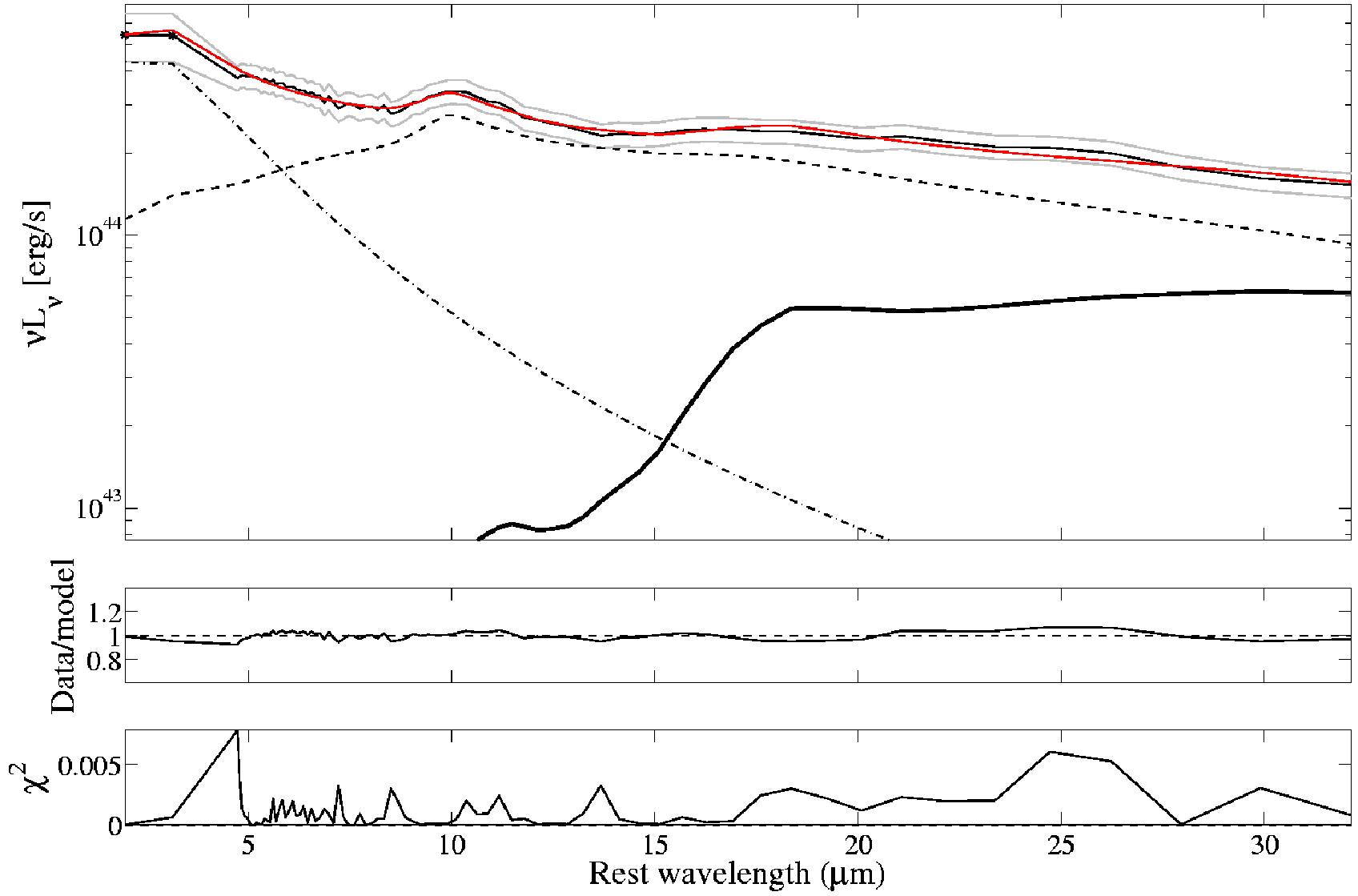}
\caption{Three component fits to the spectra of B2~2201+31A (left) and PG~1617+175 (right).
The addition of a BB component (dash-dotted line) clearly improve the quality of the fit for
these objects and the rest of our sample. Middle and bottom
panels are as in Fig.\ref{fig:1comp_fit}.}
\label{fig:3comp_fit}
\end{figure*}

Fig.~\ref{fig:3comp_fit} shows the best fit
three-component models for two representative cases, B2~2201+31A and PG~1617+175.
The top panel of each diagram shows the best fit model (in red) and the observed spectrum
(in black). We also show individual components:
torus (dashed line), NLR (thick black line) and hot BB (dash-dotted line).
Asterisks represent the K and L photometry. In the bottom and middle panels of each
diagram we show the quality of the fit in each wavelength bin.

\begin{figure}[ht] 
\plotone{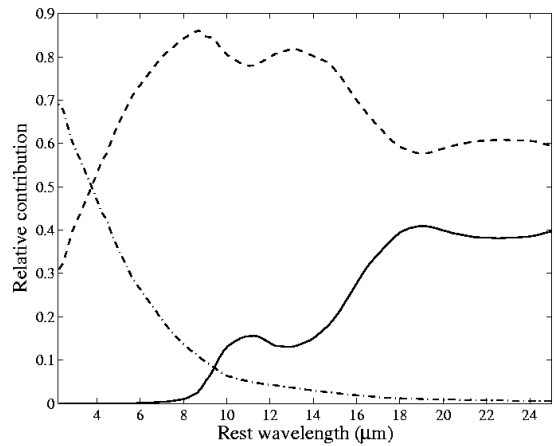}
\caption{Relative contributions of the BB component (dash-dotted line),
torus (dashed line), and NLR (solid line) to the fitted models 
(median values at each wavelength for the entire sample).}
\label{fig:relative_cont}
\end{figure}

Having obtained satisfactory fits for all sources, we have calculated the median contribution, over the entire
sample, of each of the components. This is done at every wavelength and is shown in  Fig.~\ref{fig:relative_cont} in such a way that
the  sum of all components at each wavelength bin is 1. The diagram shows that the 
 BB component dominates the spectrum below $\sim$4 \mic\ while the torus dominates
between 5-25 \mic. The NLR component has a non-negligible contribution ($\sim40$\%) above $\sim$16 \mic, including
the 18 \mic\ emission feature.

\subsection{Fit quality}
\label{sec_choose_fits}

Each of the fits is assigned a quality flag set by four criteria, the calculated \Chisq\ over
the entire 2--35 \mic\ range and partial \Chisq\ values for three shorter segments of the spectrum.
Regarding the first, since the errors are not normally distributed,
the best \Chisq\ has a different meaning from the commonly used \Chisq\
and setting a standard confidence level is not applicable.
Given this, we decided to consider only fits with \Chisq\ values not exceeding the minimum \Chisq\
by more than a factor of 1.25.  This conservative narrow range of acceptable \Chisq\ was chosen to ensure that
it contains, beside the best fit, only acceptable fits. Indeed all values within
this range seem very good and in most cases, there are also a few acceptable fits
outside of this range. However, the distribution of the model parameters for the
acceptable fits is not affected much by these outliers.

The three additional partial \Chisq\ criteria are aimed to identify those fits where a certain
part of the spectrum is not fitted well
even though the global \Chisq\ is within the above mention range. For this, we
examine the goodness of the fit in three different parts of the spectrum
by calculating \Chisq\ values for the ranges $\sim2-7$ \mic, $7-20$ \mic\ and $\sim20-35$ \mic.
We used a similar approach (in this case a factor of 1.5 relative to the best \Chisq)
and reject fits that are outside of this range.

Having fixed the \Chisq\ limits, we selected the fits that pass all
four criteria (one global and three partial \Chisq\ limits). There are thousands of possible
torus models and a much larger number of possible combinations hence,
the number of accepted fits is large, typically 10-50 per sources. All these
are inspected visually to verify the success of the automatic selection process.
The groups of acceptable models, for all sources, is the basis for the following analysis.

Two objects in our sample, PG~0157+001 and PG~0050+124, have total
covering factors larger than unity. This can perhaps be explained by source variability.
To check this for the entire sample we compare the Boroson \& Green (1992) data with more
recent observations from the SDSS. New spectra are available for 8 objects in our sample,
5 of which exhibit an increase
in luminosity by a factor $\sim2$, two show a decrease of about the same factor
and one shows no significant change.
Of the two objects with a covering factor larger than unity, only PG~0157+001
has recent SDSS spectra. A comparison with the older observations
show no significant change in luminosity
between the two epochs. Under-estimation of the star formation contribution to
the mid-infrared range of the spectrum may also result in an over estimation
of the covering factor.
Obviously, we could not fit the spectra of these two objects with adequate quality,
using the measured value of their bolometric luminosity
(see the discussion regarding the implications of \Lbol\
related uncertainties in \S\ref{sec_dis_torus_properties}).
To proceed, we assume that flux variation is the cause for the apparent discrepancy
and artificially lower the bolometric luminosity of these objects so that their covering factors become 1.
This allows us to fit their spectra in a way similar to the other sources and get adequate fits and reasonable model parameters.
However, we do not include them in the overall statistics and all parameter
distributions and mean values presented here do not include these objects.

\begin{deluxetable}{lcccccc} 
\tablecolumns{7}
\tablewidth{0pt}
\tabletypesize{\footnotesize}
\tablecaption{Torus parameters for best fit models\label{tab_torus_pars}}
\tablehead{\colhead{Object} &\colhead{$\sigma$} &\colhead{$Y$} &\colhead{$N_{0}$} &\colhead{$q$} &\colhead{$\tau_{V}$} &\colhead{$i$}\\
\colhead{} &\colhead{(deg)} &\colhead{} &\colhead{} &\colhead{} &\colhead{} &\colhead{(deg)}}
\startdata
PG 2349-014 & 29 & 35 &  7 &  1 & 80 & 48 \\ 
PG 2251+113 & 15 & 23 &  8 &  1 & 49 & 67 \\ 
PG 2214+139 & 25 & 25 &  4 &  1 & 71 & 40 \\ 
PG 1700+518 & 43 & 34 &  3 &  2 & 87 & 43 \\ 
PG 1626+554 & 28 &  6 &  6 &  1 & 17 & 31 \\ 
PG 1617+175 & 46 & 44 &  2 &  2 & 93 & 20 \\ 
PG 1613+658 & 48 & 55 &  3 &  1 & 80 & 27 \\ 
PG 1448+273 & 19 & 11 &  8 &  1 & 59 & 53 \\ 
PG 1440+356 & 40 & 12 &  4 &  0 & 41 & 26 \\ 
PG 1435-067 & 57 & 34 &  1 &  2 & 71 & 38 \\ 
PG 1426+015 & 35 & 33 &  5 &  1 & 75 & 30 \\ 
PG 1411+442 & 55 & 40 &  2 &  2 & 70 & 14 \\ 
PG 1309+355 & 22 & 24 &  6 &  1 & 67 & 37 \\ 
PG 1302-102 & 37 & 30 &  6 &  0 & 44 & 32 \\ 
PG 1244+026 & 33 & 40 &  6 &  1 & 68 & 31 \\ 
PG 1229+204 & 34 & 33 &  5 &  2 & 91 & 28 \\ 
PG 1126-041 & 34 & 36 &  6 &  1 & 61 & 28 \\ 
PG 1116+215 & 34 & 31 &  3 &  1 & 64 & 39 \\ 
PG 1004+130 & 31 & 39 &  4 &  0 & 38 & 35 \\ 
PG 1001+054 & 22 & 29 &  7 &  1 & 72 & 38 \\ 
PG 0953+414 & 18 & 20 &  7 &  1 & 66 & 35 \\ 
PG 0838+770 & 49 & 28 &  4 &  0 & 62 & 23 \\ 
PG 0026+129 & 27 & 35 &  3 &  0 & 79 & 43 \\ 
PG 0157+001 & 48 & 52 &  5 &  0 & 71 & 16 \\ 
PG 0050+124 & 51 & 35 &  5 &  0 & 20 &  8 \\ 
B2 2201+31A & 16 & 33 &  8 &  2 & 35 & 29 \\
\enddata
\end{deluxetable}

\section{Discussion}
\label{sec_discussion}

Having found the best three-component models for all individual spectra,
we now discuss the properties of these components as well as their distribution in the sample.

\subsection{Torus properties}
\label{sec_dis_torus_properties}

\begin{figure}[ht] 
\epsscale{1.1}
\vspace{0.2cm}
\plotone{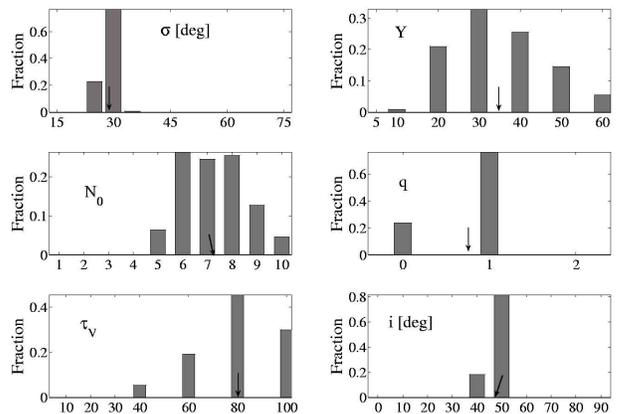}
\caption{Typical example of the distribution of acceptable torus parameters. The objects is PG~2349-014
and the arrows denote mean values. Note that $\sigma$, $q$ and $i$ have a very narrow
distribution while $N_{0}$, $Y$ and $\tau_{V}$ have a broader one.}
\label{fig:par_dist_PG2349_014}
\end{figure}

\begin{figure}[ht] 
\epsscale{1.1}
\vspace{0.2cm}
\plotone{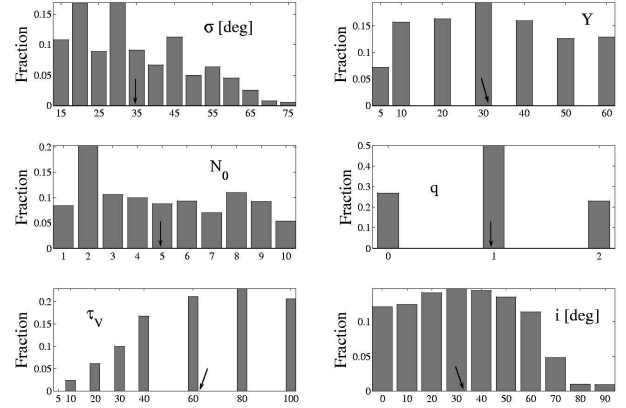}
\caption{Torus parameter distributions for the sample excluding
PG~0157+001 and PG~0050+124 (see text; arrows denote mean values).}
\label{fig:par_dist_torus}
\end{figure}

We examine the distribution of the different torus parameters for each of the objects in our sample.
Some of the parameters have a narrow distribution around a mean value and in
some cases only a single acceptable value.
Other parameters exhibit a broader, more uniform distribution.
In Fig.~\ref{fig:par_dist_PG2349_014} we present a typical example of the acceptable torus parameters for
one source, PG~2349-014. As seen in this case, $\sigma$, $i$, and $q$ have very narrow distributions for all
acceptable torus models. $N_{0}$ ,$Y$, and \tv\ have 
broader distributions around their mean values. The values obtained for individual objects
from their best fit models are listed in table~\ref{tab_torus_pars}
The parameter distributions of all sources were combined by giving each value within the
acceptable range in a certain source its relative weight in the distribution.
The results of this combination are general parameter distributions
that are shown in Fig~\ref{fig:par_dist_torus}.
Inspection of the various parts of this figure lead to the following conclusions:
\begin{enumerate}
\item
The mean torus width parameter is $\sigma=34^{\circ}$.
None of the fits requires the largest allowed $\sigma$ of $75^{\circ}$.
This value determines, more than any of the other parameters, the geometrical covering factor ($f_2$)
of the torus and is consistent with the expected type-I:type-II ratio (see below).
\item
The mean radial extent of the torus, Y, has a broad distribution with a mean value of 31.
The range in Y implies torus outer sizes, which range from $\sim$1 to 35 pc using $R_{d,C}$
and $\sim$3 to 90 pc using $R_{d,Si}$.
\item
The average number of clouds along an equatorial ray, $N_{0}$,
has a broad distribution with a mean of 5 clouds.
The distribution is practically uniform for $N_{0}\geq3$.
\item
The parameter that specifies the radial power-law distribution of the clouds, q,
has a mean value of 1. This parameter is related to the anisotropy of the torus radiation.
As q decreases the torus radiation becomes less isotropic. The SED shape is affected less
by q and hence we do not attach great significance to its specific value.
\item
The mean optical depth of a cloud is \tv=58. This parameter, again, has a broad distribution
covering the range \tv$\geq20$. Since the requirement for large MIR optical depth is
\tv$\sim 10$, it is not surprising that the torus models are not
very sensitive to the exact value of \tv\ beyond this value.
\item
The torus inclination angle shows a broad distribution for all $i \leq 60^{\circ}$ with a mean of
$33^{\circ}$. This again is consistent with the assumption that the direct line-of-sight
to the center of type-I AGNs is almost completely free of obscuring material.
\end{enumerate}

We also checked for various correlations between those parameters and found no such correlation.
This suggest that torus models with different parameters can result in similar SEDs.

As explained, SED related uncertainties, in particular the bolometric luminosity of
the sources, make it difficult to determine the exact torus properties in individual
objects. There is a relatively large range in some of the parameters (e.g. $N_0$ and $i$)
and small changes in \Lbol\ can result in a significant change in those parameters.
For example, lower bolometric luminosity, due to changes in \Lop\
between the optical and the \spitzer observations, requires lower $L_{\rm \nu,torus}$
and higher inclination angle. In principle this may introduce non-negligible uncertainties
on $i$ (and other parameters) in individual sources.
However, the average sample properties
are less likely to depend on such uncertainties and the discussion below focus on this aspect
of the work.
Below we examine three of the parameters, $i$, $N_0$, and $\sigma$ using Eq.~\ref{eq:Pesc}.
We set two of the parameters to their mean values and investigate the dependence of
$P_{\rm esc}$ on variations of the remaining free parameter.
This is equivalent to the investigation of a generic torus that represents the entire
sample of 26 AGNs.

\begin{deluxetable*}{lcccccc} 
\tablecolumns{7}
\tablewidth{0pt}
\tabletypesize{\footnotesize}
\tablecaption{Correlation p values\label{tab_corr_coef}}
\tablehead{
&
\multicolumn{3}{c}{Pearson} &
\multicolumn{3}{c}{Spearman}
\\
\\Parameter
&
\colhead{$log(L_{\rm bol})$} &
\colhead{$log(M_{\rm BH})$} &
\colhead{$L/L_{\rm Edd}$} &
\colhead{$log(L_{\rm bol})$} &
\colhead{$log(M_{\rm BH})$} &
\colhead{$L/L_{\rm Edd}$}
}

\startdata
$\sigma$ & 0.031(-) & ... & ... & 0.028(-) & ... & ... \\
$Y$ & ... & ... & ... & ... & ... & ... \\ 
$N_{0}$ & ... & ... & ... & ... & ... & ... \\ 
$q$ & ... & ... & ... & ... & ... & ... \\ 
$\tau_{V}$ & ... & ... & ... & ... & ... & ... \\ 
$i$ & 0.014(+) & ... & ... & 0.0047(+) & ... & ... \\ 
$P_{\rm esc}$ & ... & ... & ... & ... & ... & ... \\ 
$\RNLR$ & 0.00018(+) & ... & 0.0054(+) & 0.0004(+) & ... & ... \\
$\RNLR / R_{d}$ & ... & ... & 0.01(+) & ... & ... & ... \\
BB temperature & ... & ... & ... & ... & ... & ... \\ 
Real total CF & 0.0033(-) & 0.024(-) &  & 0.0056(-) & 0.045(-) & ... \\ 
Apparent total CF & 0.0011(-) & 0.0063(-) &  & 0.0013(-) & 0.014(-) & ... \\ 
$f_{2}$ & 0.026(-) & ... & ... & 0.029(-) & ... & ... \\ 
$f(i)$ & 0.0051(-) & 0.026(-) & ... & 0.0026(-) & 0.016(-) & ... \\
NLR CF & ... & ... & ... & ... & ... & ... \\ 
BB CF & 0.013(-) & ... & ... & 0.04(-) & ... & ... \\ 
$f_{2}$+BB CF & 0.0046(-) & 0.048(-) & ... & 0.0045(-) & ... & ... \\
$M_{\rm torus}$ & 0.013(+) & 0.019(+) & ... & 1.7e-06(+) & 0.0003(+) & ... \\
$M_{\rm torus}/M_{\rm BH}$ & ... & ... & ... & ... & ... & 0.0083(+) \\ 
$M_{\rm torus}/L_{\rm bol}$ & ... & ... & ... & ... & ... & ... \\ 
\enddata
\tablecomments{Plus and minus signs indicate positive and negative correlations, respectively.
Results for $p>0.05$ are not listed.}
\end{deluxetable*}

First we consider the torus inclination angle $i$ as the free parameter.
Using Eq.\ref{eq:Pesc} together with the mean values found here, the probability
that light from the central region will escape such a structure, and the source will be
classified as type-I AGN, is about 75\%. Setting $N_{0}$ and $\sigma$ to their mean values
(5 and $34^{\circ}$, respectively) and
changing the inclination angle, we find that for
$i= 50^{\circ}$ $P_{\rm esc}\simeq30\%$ and for $i=70^{\circ}$ $P_{\rm esc} < 3\%$.
We therefore expect that for type-I objects, the inclination angle should not drop below
 $\sim60^{\circ}$. As can be seen in Fig.\ref{fig:par_dist_torus},
the distribution of the inclination angle is indeed consistent with random inclination angles
in the range $0-60^{\circ}$ with no preference to a specific angle.

The second important parameter is the mean number of clouds, $N_0$.
Setting $\sigma$ and $i$ to their mean values and
changing $N_{0}$, we find that $P_{\rm esc}$ remains large even for very high values of $N_{0}$.
Thus, the exact value of $N_{0}$ is not very important provided the inclination angle
is not too large. This is the reason why $N_{0}$ has the broad distribution seen in
Fig.\ref{fig:par_dist_torus}.

Next we tested the acceptable range of the torus width parameter, $\sigma$, by fixing all
other parameters to their mean values.
We find that $P_{\rm esc}$ is sensitive to such changes and falls rapidly for $\sigma\gtrsim45^{\circ}$.
This is the reason that the $\sigma$ distribution falls rapidly for $\sigma\gtrsim45^{\circ}$.
Torus with high value of $\sigma$ would have low $P_{\rm esc}$ and thus
a low probability of being classified as type-I AGN even if the inclination angle
is zero (a face-on torus).

The combined effect of the above parameters is, perhaps, a more meaningful test.
In particular, the combination of
$\sigma$ and $N_{0}$ determine the geometrical covering factor of the torus, $f_2$,
which is combined, later, with the equivalent property of the other components.
This issue is discussed in \S\ref{sec_dis_est_CF}.

The above procedure is not applicable to the visual optical depth parameter.
This parameter does not affect $P_{\rm esc}$ or $f_2$ but influence
the SED shape, in particular the apparent strength of the silicate emission features.
The mean visual optical depth for the entire sample is \tv=58 and the lowest value is $20$.
Thus, realistic clumpy torus models require large IR
optical depths to explain the observed spectrum. The exact value of \tv\
is not very important once the clouds are thick enough.

The torus size found here is consistent with upper limits set by several high resolution observations
of nearby AGNs (e.g, Soifer et al. 2003; Radomski et al. 2003; Jaffe et al. 2004).
This is in contrast to previous theoretical works involving smooth density distribution torus models.
In these models, the torus is required to be much larger (up to few hundreds of pc)
in order to posses the large amounts of cool dust necessary for producing the observed MIR emission.
This requirement arises because in smooth density distribution models, the dust temperature
is uniquely determined by the distance from the center. In contrast, the clumpy torus model
enables emission from cool dust clouds that are much closer to the center thus able of producing the required MIR
emission with much smaller dimensions.

Perhaps the more interesting results are the correlations of the torus
parameters with the physical properties of the AGNs,
in particular BH mass (\mbh), bolometric luminosity and normalized accretion rate, \Ledd.
To study these correlations we have estimated these quantities for all sources using the procedure described in Netzer \& Trakhtenbrot (2007). In this procedure (the ``virial" mass determination, see Eq.~1 in Netzer \& Trakhtenbrot 2007), \Lop\ and FWHM(\Hb) are combined to obtain \mbh, and \Ledd\, is obtained by using the assumed bolometric correction factor BC.
Table~\ref{tab_corr_coef} lists p-values for Pearson and Spearman-rank correlation tests
for the different torus parameters against those properties.
These values represent the probability that the observed correlation occurred by chance.
We regard as significant all correlations where $ p < 0.05$  and indicate with plus and minus signs
positive and negative correlations, respectively.
Out of the above geometrical parameters, only $\sigma$ and $i$ show a significant correlation with \Lbol.
This is consistent with a 'receding torus' assumption (Lawrence 1991) and discussed in \S\ref{sec_dis_est_CF}.

\subsection{NLR properties}
\label{sec_dis_NLR_properties}

The important parameters of the NLR component are its distance from the center, which determines
the dust temperature, and its covering factor, which determines the fraction of
MIR flux emitted by the clouds ($a_{\rm NLR}$ in Eq.\ref{eq:Fmodel}).

The photoionization models used to calculate the NLR emission are all of the same
constant density ($n_{\rm H}=10^{5}\cc$) and column density ($10^{21.5}$ \cmii).
As explained earlier, the clouds are optically thin in the MIR range and their distance from the
central source uniquely determine the NLR dust temperature. The assumed gas density plays no important role
and serves only as a convenient way to determine the photoionization model parameters (in the actual
calculations, the model
parameters are density and ionization parameter).
The dust column density is determined by the assumed gas composition and depletion.
The present calculations use solar composition with abundances smaller than assumed in several recent NLR models
such as the constant pressure models of Groves et al. (2006). This again is of little importance
since the dust column density is the only important parameter, unless an unusual grain mixture
is used.

\begin{figure} 
\epsscale{1.1}
\plotone{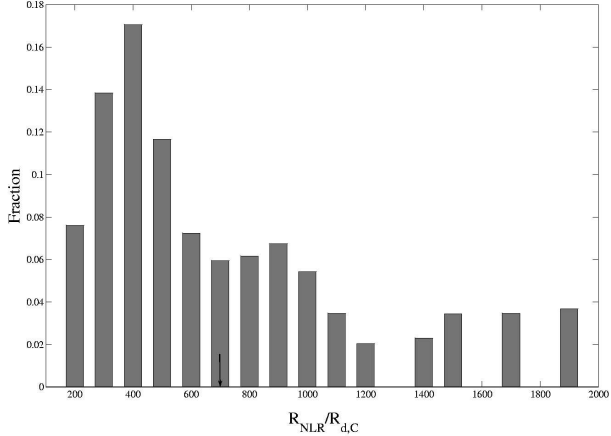}
\caption{NLR distance distribution in units of the graphite sublimation radius.}
\label{fig:NLR_dist}
\end{figure}

Table \ref{tab_NLR_prop} lists NLR distances for
the 26 best models. Radii are given in pc as well as in units
of the sublimation distances, $R_{d,C}$ and $R_{d,Si}$.
For the graphite sublimation radius, the NLR distances
span the range $256 - 1755 R_{d,C}$ with a mean value of 707$R_{d,C}$.
This translates to $101 - 694 R_{d,Si}$ with a mean value of 280$R_{d,Si}$.
The distribution of \RNLR in units of the graphite sublimation
radius is shown in figure~\ref{fig:NLR_dist}.

The mean distance found here is about a factor of 4 larger than the values found by S08.
The reason for the larger distances is mostly due to the introduction of real
torus models compared with the combination of BBs used in S08. The clumpy torus
models contribute a certain amount to the observed silicate emission
and hence the NLR contribution, which was the only one with such features considered in S08, is reduced.
This can also be seen when comparing the NLR covering
factor found here and the one found in S08 (see Table~\ref{tab_modeling_components_CF}).
The mean covering factor for the NLR component in S08 is a factor 2 larger
than the one found here which is $\sim$7\%.
We consider the present values more appropriate to the objects in question (see \S\ref{sec_dis_est_CF})

\begin{figure} 
\plotone{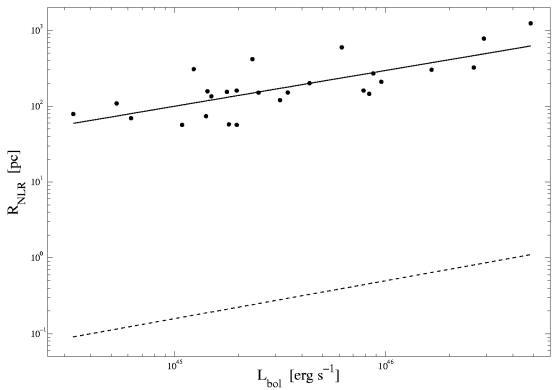}
\epsscale{1.1}
\caption{NLR cloud distances against $L_{\rm bol}$.
Circles represent the NLR distances in pc, and the solid line is a
linear fit to the data with a slope of 0.47.
The graphite dust sublimation radius is shown for comparison as a dashed line.}
\label{fig:Lbol_vs_RNLR}
\end{figure}

Figure~\ref{fig:Lbol_vs_RNLR} shows NLR cloud distances against the
bolometric luminosity of the objects.
Using the values in Table \ref{tab_NLR_prop} we find the following scaling relationship,
\begin{equation}
\RNLR = 295 \times L_{46}^{0.47\pm0.13} \,\,  {\rm pc}.
\label{eq:R-NLR_Lbol_excatfit}
\end{equation}
The slope is consistent with 0.5, which is the one predicted for simple single-cloud photoionization
models and a constant $U$ gas.

\begin{deluxetable}{lcccc} 
\tablecolumns{5}
\tablewidth{0pt}
\tabletypesize{\footnotesize}
\tablecaption{NLR properties for best fit models\label{tab_NLR_prop}}
\tablehead{\colhead{Object} &\colhead{$\log(U)$\tablenotemark{a}} &\colhead{\RNLR} &\colhead{$\RNLR/R_{d,C}$} &\colhead{$\RNLR/R_{d,Si}$}\\
           \colhead{} &\colhead{} &\colhead{(pc)} &\colhead{} &\colhead{}}
\startdata
PG 2349-014 & -2.6 & 201 & 608 & 240\\
PG 2251+113 & -2.2 & 323 & 400 & 158\\ 
PG 2214+139 & -2.0 & 58 & 271 & 107\\ 
PG 1700+518 & -2.9 & 780 & 913 & 361\\ 
PG 1626+554 & -1.6 & 161 & 726 & 287\\ 
PG 1617+175 & -2.8 & 157 & 832 & 329\\
PG 1613+658 & -1.6 & 152 & 518 & 205\\ 
PG 1448+273 & -2.2 & 74 & 392 & 155\\ 
PG 1440+356 & -3.5 & 308 & 1755 & 694\\ 
PG 1435-067 & -2.7 & 155 & 737 &  291\\ 
PG 1426+015 & -1.8 & 57 & 256 & 101\\ 
PG 1411+442 & -2.7 & 135 & 701 & 277\\ 
PG 1309+355 & -2.1 & 161 & 363 & 144\\ 
PG 1302-102 & -2.2 & 210 & 431 & 170\\ 
PG 1244+026 & -2.8 & 79 & 867 &  343\\
PG 1229+204 & -2.5 & 69 & 558 & 221 \\ 
PG 1126-041 & -3.0 & 109 & 946 & 374\\ 
PG 1116+215 & -2.6 & 270 & 579 & 229\\ 
PG 1004+130 & -2.3 & 302 & 470 & 186\\ 
PG 1001+054 & -2.5 & 151 & 603 & 239\\ 
PG 0953+414 & -2.0 & 145 & 318 & 126\\ 
PG 0838+770 & -2.0 & 57 & 345 & 136\\ 
PG 0026+129 & -2.4 & 120 & 428 & 169\\ 
PG 0157+001 & -3.4 & 597 & 1520 & 601\\ 
PG 0050+124 & -3.6 & 417 & 1728 & 683\\ 
B2 2201+31A & -3.1 & 1234 & 1121 & 443\\
\hline \\
Mean 	    &      & 249  & 707	 & 280
\enddata
\tablenotetext{a}{Assuming $n=10^{5}\, \cc$}
\end{deluxetable}

The dusty NLR dimensions can be compared with earlier
works on NLR size such as Bennert et al. (2002), Schmitt et al. (2003),
Netzer et al. (2004) and Baskin and Laor (2005).
These papers use two different methods to estimate the NLR size.
The Bennert et al. and Schmitt et al. works are based on detailed HST imaging of AGN samples
that are of the same redshift range as the one considered here.
Our dust temperature-based dimensions are smaller by a factor of $\sim10$ compared
with these imaging-based estimates.
This is not surprising since our estimates are biased toward those NLR clouds that have the
largest covering factor for the hottest dust clouds.
The discrepancy suggests that real NLRs, i.e. those with a range of dust temperatures
contribute more to the longer wavelength part of the spectrum.
Furthermore, Netzer et al. (2004) suggested that the
Bennert et al. (2002) estimates are considerably larger than
``real" NLR sizes for two reasons. The first is that the
$\RNLR\propto L_{\rm ion}^{1/2}$ dependence must break down
at some intermediate luminosity since extrapolating this relation to high
luminosity objects would result in NLR dimensions that exceed the host galaxy dimension.
The second reason is that Bennert et al. (2002) measured image sizes that
encompasses more than 98\% of the detectable \oiii\ emission.
A 90\% or 95\% intensity limits can lead to much smaller dimensions.

Baskin \& Laor (2005) used a different method to estimate
the distances to 40 NLRs in the Boroson \& Green (1992) sample.
They fitted the observed intensities of the \Oxs, \Oxf\ and \Hb\ narrow emission lines using two different models.
The first, assumed a single uniform, \Oxs\ emitting region.
This resulted in $\RNLR \approx 130 \times L_{46}^{0.45} \rm{pc}$
\footnote{In this expression and the following ones, we converted the continuum luminosity
given by Baskin \& Laor (2005) into bolometric luminosity.}
and a very large mean gas density of $10^{5.85} \cc$.
These dimensions are smaller than those found by Bennert et al. (2002).
The second model assumed two emitting regions:
a low, constant gas density ($n=10^{3}\,\cc$) region where most of the \Oxs\ originates and
a high, constant gas density ($n=10^{7}\,\cc$) region where most of the \Oxf\ originates.
For the low density region they found $\RNLR \approx 1850 \times L_{46}^{0.34} \rm{pc}$
and for the high density region $\RNLR \approx 4 \times L_{46}^{0.5} \rm{pc}$.
The \Oxs\ dimensions are comparable to those of Bennert et al. (2002).
It is clear that single-zone and even two-zone NLR models are highly simplified.
Moreover, dimension deduced from a narrow line intensity is uncertain since the gas density is
changing across the NLR and in general $\RNLR\propto n^{-0.5}$. Thus, it is not
surprising that the NLR size obtained from such estimates spans a very large range.

The M82 starburst template subtracted from the spectra could, in principle, have an effect
on the calculated NLR properties, e.g. a different template could have larger contribution
towards shorter wavelengths ($\sim$30 \mic). Consequently, the NLR component
would have less weight and different spectral shape that would result in larger NLR distances.
We regard this possibility as an additional uncertainty on the determination of the NLR distance.

In conclusion, our hot dust measure of the NLR size is based on a method
different than the direct imaging method and the photoionization modelling method.
While further discussion of those discrepancies is beyond
the scope of the present paper, the main conclusion is that simple, dusty NLR clouds (or single NLR shells)
at the distances found above, in combination with the two other model components, can
adequately fit the 2--35 \mic\ spectrum of the QUEST AGNs.

\subsection{Hot BB properties}
\label{sec_dis_BB_properties}

\begin{figure} 
\epsscale{1.1}
\vspace{0.2cm}
\plotone{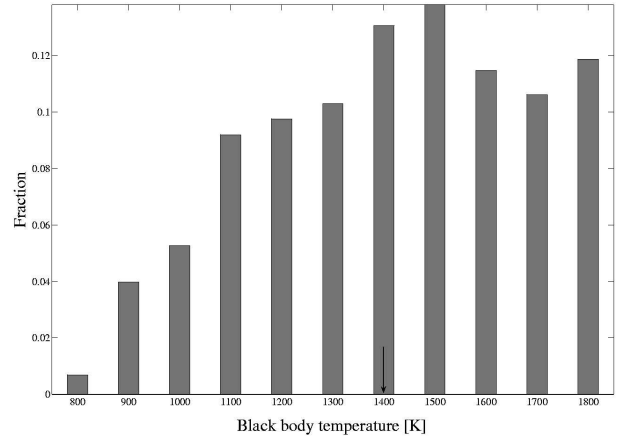}
\caption{Blackbody temperature distribution for the hot dust component.}
\label{fig:BB_temp_dist}
\end{figure}

The $2-4 \mu m$ wavelength range is dominated by the hot BB component of our model (see Fig.~\ref{fig:relative_cont}).
On average, the luminosity of this component is about 40\% of the total
$2-35\mu m$ luminosity and is comparable to the luminosity radiated by the clumpy torus.
The mean BB temperature is 1400 K and the distribution of temperatures
is shown in Fig.~\ref{fig:BB_temp_dist}. 
While such a large contribution is evident in all of the sources,
the physical origin of this component is not yet clear.
Clumpy tori of the type considered here cannot produce more than one luminosity
bump. Since most of the torus radiation peaks at 5--20 \mic, the short wavelength excess cannot result
from this structure. The conclusion is that the assumed BB must be a separate component.

We have considered several possible explanations for the hot BB component.
The first is a contribution
from old stellar population near the galactic center.
To examine this, we use \mbh\ to estimate the host galaxy luminosity
using the scaling relations of Lauer et al. (2007; see their Eq. 3).
To convert visual magnitudes to K--band magnitudes we use
(V-K)=3.2 typical of giant elliptical galaxies (e.g. Grasdalen 1980).
This K--band luminosity was compared to the luminosity of the
BB found here.
In all cases, the BB component is about a factor 10 more luminous than the derived
K--band luminosity of the host galaxy. Thus, such a stellar population would not have
sufficient luminosity to account for the observed 2--7 \mic\ flux.

The space distribution of such an old stellar population is another limitation. Broad band monitoring
of several nearby, intermediate luminosity AGN clearly show K-band variations on time scales of
weeks to months (e.g., Suganuma et al. 2006). This has been
interpreted as changes in the temperature of the innermost dust, just beyond the sublimation radius.
Given the great similarity between the 2--35 \mic\ spectrum of the
present sample and the spectra of those variable AGNs, we conclude that the origin
of the K and the L-band emission, in the present sample and the above AGN sample, is the same.
Such dimensions are clearly too small for the assumed stellar population.
Needless to say, any K or L-band variations are inconsistent with stellar population properties.

Another strong source of radiation is starburst activity in the host galaxy.
The IR emission from starburst, however,
is expected to peak at much larger wavelengths.
S06 found a strong correlation between the strength of the PAH feature at 7.7 \mic\ and the continuum luminosity
at 60 \mic. The same trend is found in starburst dominated ultraluminous infrared galaxies (ULIRGs; see Fig.4 in S08).
Thus, luminosity produced by starburst in the QUEST sample dominates the far-IR part of the spectrum and
cannot account for the hot dust component.

Another possibility that has already been mentioned is that the new component represents
emission by very hot dust, which is not a part of the clumpy torus structure.
As explained in \S\ref{sec_res_3_components}, we could not fit the observations
assuming single hot dust clouds of the same grain composition as the torus.
The reason is that such clouds produce extremely strong 10 \mic\ emission.
This is not the case for pure graphite dust that has no prominent features at MIR
wavelengths. Graphite dust grains can survive at the BB
temperatures found here and provide a suitable explanation for the BB component.
The location of this dust can be inside the silicate sublimation radius but still
outside the dust-free broad line region.
A full discussion of this possibility, including realistic hot graphite
grains spectra, is beyond the scope of this paper.

Finally we note that the above suggested location, close or even inside the silicate dust
sublimation radius, is also problematic from the point of view of the derived
covering factors. Our calculations assume no obscuration of one component
by another yet such a location must obscure some of the radiation impinging on the torus.
The location of these clouds imply that the radiation incident upon the torus
should change and include part of the re-emitted radiation from the BB component.
Given the unknown properties of the hot BB component, we cannot take this into account
within the framework of the present work.

\subsection{Covering factors}
\label{sec_dis_est_CF}

\begin{figure} 
\epsscale{1.1}
\vspace{0.2cm}
\plotone{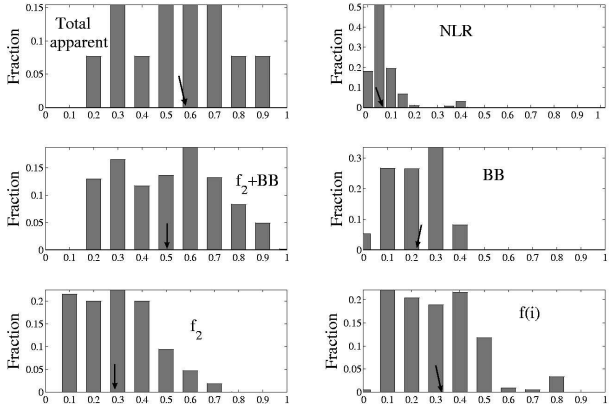}
\caption{Distributions of different covering factors (arrows denote mean values).
The total apparent CF is the the ratio between the integrated MIR luminosity
and the bolometric luminosity.
$f(i)$ is the ratio between the integrated, angle dependent, torus MIR luminosity and the bolometric
luminosity of the source. This differs slightly from $f_2$ due to the anisotropic radiation of the torus.
Note that PG~0157+001 and PG~0050+124 have been omitted since their CF exceed unity (see text).}
\label{fig:CF_All}
\end{figure}

A major assumption of this work is that the entire MIR spectrum, after
starburst subtraction, is reprocessed AGN radiation. This can be used
to deduce the covering factor of the central source by the three components
(see also S08 and references therein; Maiolino et al. 2007).
Regarding the BB and the NLR components, this factor is obtained directly from the comparison of their
deduced luminosities with \Lbol.
The covering factor of the torus, $f_2$,
is different because of its non-isotropic radiation.
For a given model, $f_2$ is calculated from the torus parameters.
To demonstrate these differences, we compare in Table~\ref{tab_modeling_components_CF},
$f_2$, $f(i)$, and the covering factors of the NLR and the BB components.
The total covering factor can be defined in two ways:
\textbf{1.} Apparent total covering factor,
defined as $L_{\rm obs}(2-100)/L_{\rm bol}$. This is similar to the number used in all earlier investigations of this type.
Since the \textit{Spitzer}-IRS spectra extend only to $\sim$35 \mic, we use the mean ``intrinsic"
AGN SED of Netzer et al. (2007), obtained from the starburst subtracted QUEST spectra, to calculate the ratio between $L_{\rm obs}(2-100)$ and $L_{\rm obs}(2-35)$.
This ratio is 1.072 and thus we multiply $L_{\rm obs}(2-35)$ of each object by this factor to get its apparent total covering factor.
\textbf{2.} Real total covering factor, calculated by summing together $f_2$ and the NLR and BB covering factors.
The various covering factor distributions are shown in Fig.~\ref{fig:CF_All}.

The mean value of $f_2$ is 0.27. This value
is lower than the $f(i)$, for which
the mean value is 0.34. The difference is due to the anisotropy
of the torus radiation and the fact that, for the present sample of type-I sources,
the torus inclination angle is small.
The mean covering factor of the NLR component is 0.07. This is in agreement
with other estimates that are based on narrow emission line imaging and line
equivalent width measurements, which suggest NLR covering factor
$<10$\% (e.g., Baskin \& Laor 2005; Netzer \& Laor 1993).
The mean covering factor of the BB component is 0.23.
The mean real total covering factor in our sample is 0.57, which is slightly lower
than the mean apparent total covering factor of 0.59.

\begin{figure} 
\epsscale{1.1}
\vspace{0.2cm}
\plotone{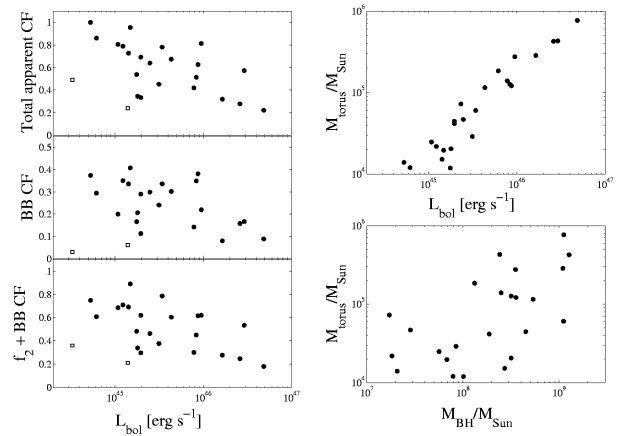}
\caption{Correlations between covering factors and bolometric luminosities (left panels)
and between torus mass and bolometric luminosity and the mass of the black hole (right panels).
The square symbols represent sources with the lowest \mbh\ in the sample, see text for explanation}
\label{fig:corrs_Lbol}
\end{figure}

\begin{deluxetable*}{lcccccc} 
\tablecolumns{7}
\tablewidth{0pt}
\tabletypesize{\footnotesize}
\tablecaption{Covering Factors\label{tab_modeling_components_CF}}
\tablehead{
\colhead{Object} &\colhead{Real total CF} &\colhead{Apparent total CF}
	 &\colhead{Real torus CF ($f_2$)} &\colhead{$f(i)$} &\colhead{NLR CF} &\colhead{BB CF}}
\startdata
PG 2349-014 & 0.70 & 0.67 & 0.30 & 0.38 & 0.09 & 0.30 \\ 
PG 2251+113 & 0.30 & 0.28 & 0.09 & 0.08 & 0.05 & 0.16 \\ 
PG 2214+139 & 0.36 & 0.34 & 0.13 & 0.15 & 0.03 & 0.21 \\ 
PG 1700+518 & 0.58 & 0.57 & 0.37 & 0.42 & 0.05 & 0.17 \\ 
PG 1626+554 & 0.30 & 0.33 & 0.18 & 0.21 & 0.00 & 0.11 \\ 
PG 1617+175 & 0.72 & 0.73 & 0.36 & 0.44 & 0.03 & 0.34 \\ 
PG 1613+658 & 0.81 & 0.78 & 0.45 & 0.60 & 0.02 & 0.34 \\ 
PG 1448+273 & 0.24 & 0.24 & 0.15 & 0.16 & 0.03 & 0.06 \\ 
PG 1440+356 & 0.74 & 0.79 & 0.36 & 0.45 & 0.04 & 0.35 \\ 
PG 1435-067 & 0.55 & 0.54 & 0.32 & 0.34 & 0.07 & 0.17 \\ 
PG 1426+015 & 0.68 & 0.69 & 0.33 & 0.46 & 0.06 & 0.29 \\ 
PG 1411+442 & 0.92 & 0.95 & 0.48 & 0.59 & 0.03 & 0.41 \\ 
PG 1309+355 & 0.40 & 0.42 & 0.16 & 0.20 & 0.10 & 0.14 \\ 
PG 1302-102 & 0.77 & 0.81 & 0.40 & 0.56 & 0.14 & 0.22 \\ 
PG 1244+026 & 0.44 & 0.49 & 0.33 & 0.46 & 0.08 & 0.03 \\ 
PG 1229+204 & 0.74 & 0.86 & 0.31 & 0.41 & 0.13 & 0.29 \\ 
PG 1126-041 & 0.91 & 1.00 & 0.37 & 0.53 & 0.16 & 0.37 \\ 
PG 1116+215 & 0.65 & 0.63 & 0.23 & 0.28 & 0.04 & 0.38 \\ 
PG 1004+130 & 0.33 & 0.32 & 0.20 & 0.24 & 0.05 & 0.08 \\ 
PG 1001+054 & 0.58 & 0.64 & 0.16 & 0.21 & 0.12 & 0.30 \\ 
PG 0953+414 & 0.51 & 0.51 & 0.10 & 0.12 & 0.06 & 0.35 \\ 
PG 0838+770 & 0.77 & 0.80 & 0.48 & 0.65 & 0.09 & 0.20 \\ 
PG 0026+129 & 0.45 & 0.45 & 0.14 & 0.16 & 0.07 & 0.24 \\ 
PG 0157+001 & 1.05 & 1.08 & 0.53 & 0.80 & 0.42 & 0.11 \\ 
PG 0050+124 & 0.89 & 1.09 & 0.63 & 0.90 & 0.03 & 0.23 \\ 
B2 2201+31A & 0.20 & 0.22 & 0.09 & 0.12 & 0.02 & 0.09 \\
\hline \\
Mean & 0.57 & 0.59 & 0.27 & 0.34 & 0.07 & 0.23 
\enddata
\end{deluxetable*}

The dependence of the different covering factors on bolometric luminosity
is seen in Fig.~\ref{fig:corrs_Lbol}. Looking at the entire sample, we find no significant
correlation of bolometric luminosity with any of the covering factors. However, two of the
sources, PG~1448+273 and PG~1244+026, seem to be affecting the entire correlation in a way that
their removal from the sample makes the luminosity CF correlations significant.
These sources are marked with different symbols on the diagram.
The reason why these objects should be omitted from the sample is not clear. We note, however, that these are the sources
with the lowest \mbh\ thus may represent different properties. In particular, they are likely to be
the highest amplitude variables and the highest accretion rate BHs thus their derived \Lbol\ may be wrong.
Visual inspection of the bottom left panel shows that, without the two sources, the geometrical covering factor ($f_2$+BB) decreases
with \Lbol\ and its highest value, at the low luminosity end, is roughly 0.7, in agreement with typical assumed
distributions between type-I and type-II AGN in the local universe.

Several earlier studies found clear anti-correlation between covering factor and \Lbol\
(e.g., Wang et al. 2005; Richards et al. 2006; Maiolino et al. 2007; Treister et al. 2008).
The decrease in covering factor translates, within the unification scheme, to
a decreasing fraction of obscured AGNs as a function of luminosity.
All earlier studies were based on the total MIR emission and took no account of the various
contributers to the flux and the fact that the geometrical covering factor cannot be obtained, directly, from L(IR)/\Lbol.
Our results are consistent with these studies but take into account, more
accurately, the exact torus geometry and the needed translation between observed
IR flux and geometrical structure. They show that $\sim7\%$ of the derived
covering factor is due to NLR clouds and properly account for the inclination
angle distribution within the clumpy torus scenario.
Taking the values found here we conclude that for \Lbol$\simeq 5\times10^{45}\,\ergs$
there is roughly 1:1 ratio between type-I and type-II sources
(note that the NLR contribution is subtracted from the total covering factor).
Both Maiolino et al. (2007) and Treister et al. (2008) find a somewhat larger value for this bolometric
luminosity. However, estimates of this ratio based on X-ray surveys are similar to the one found here
and consistent across the entire luminosity range in our sample
(e.g., Hasinger et al. 2004; Treister \& Urry 2006).

The physical mechanism responsible for the decrease of covering factor with \Lbol\ is still undetermined.
One possibility is a receding torus mentioned above, where higher luminosity implies larger
dust sublimation distance, and hence an obscuring structure that is located farther away from the center.
In this scenario, the obscuring structure must have a constant height.
In the clumpy torus model this scenario corresponds to a decrease of $\sigma$ with \Lbol.
Indeed, we find that $\sigma$ is anti-correlated with the bolometric luminosity.

We also find a positive correlation between the inclination angle of the torus and the bolometric luminosity.
This is also consistent with a receding torus. An obscuring structure located at a larger distance, in this
scenario, implies a larger solid angle through which an unobscured view to the center is possible.
Thus, the luminosity of the source determines the possible range of inclination angles in which
the model would be consistent with a type-I AGN SED. Since there is no preference to a specific angle
within that range, the correlation is due to the expansion of the possible range in $i$ with increasing luminosity.

\subsection{Torus mass}
\label{sec_dis_torus_mass}
Given the torus parameters, we can estimate the torus mass from
\begin{equation}
\frac{M_{torus}}{\Msun} = 10^{4}  \, \sin(\sigma) \, N_{0} \, N_{H,23} \, \left(\frac{R_{d}}{\rm pc}\right) ^2 \, Y \, I_{q}(Y)\,\, ,
\label{eq:torus_mass}
\end{equation}
where $N_{\rm H,23}$ is the column density of a single cloud in units of
$10^{23}$ \cmii\ obtained from \tv\ and the assumed dust-to-gas ratio.
$I_{q}$ is a function of Y and has the values
of 1, $Y/(2lnY)$ and $\frac{1}{3}Y$ for $q=$ 1, 2, and 0, respectively.

The torus mass shows a clear correlation with both the bolometric
luminosity and the mass of the central black hole (see Fig.~\ref{fig:corrs_Lbol}).
These correlations are expected since the bolometric
luminosity appears in the torus mass calculation and, in general, $L_{\rm bol}\propto$\mbh.
No correlation is found between the torus mass and \Ledd.
We also checked for correlation between $M_{\rm torus}/L_{\rm bol}$
and \Lbol\ and found none.

The mass of the torus ranges from $8.45\times10^{4}$ to $3.12\times10^{7} \Msun$ for the silicate dust
sublimation radius.
The ratio between the torus and the BH masses is ranges from $4.5\times10^{-4}$ to 0.08 and does not show any correlation with the source luminosity.

Assuming a mass-independent BH accretion efficiency of $\eta = 0.1$,
we can calculate the time it would take for the torus mass to
be completely accreted by the BH.
This is an extremely short duration of about $\sim10^{5}\,\rm{yr}$.
This time scale is similar for all sources
and is independent of torus mass, source luminosity and
any of the other properties.
The typical e-folding time for a BH mass of 10$^8$\Msun\ and
the above $\eta$ is about $4\times10^7(L/L_{\rm Edd})^{-1} \,\rm{yr}$.
This mean that, if all accretion to the center goes through the torus,
there are more than 100 short accretion episodes per one e-folding growth time.

The above result indicates at least two different possibilities:
\textbf{1.} A constant replenishment of the torus material
by larger scale mass inflow through the plane of the torus.
Such a scenario, requires a constantly changing torus structure.
\textbf{2.} Matter accreting onto the BH does not originate in the torus.
Such a scenario has been suggested by Elitzur \& Shlosman (2006), who assumed that mass from
the host galaxy arrives directly into the accretion disk.
Consequently, wind from the accretion disk forms the toroidal obscuration. As long as the
large scale mass infall is sufficiently large and continuous, the wind sustains a steady outflowing
clumpy structure, which is the clumpy torus discussed in this work.
In this case, the small mass of the torus is not directly related to the growth time of the BH.

\begin{acknowledgements}
We are grateful to Dieter Lutz, Mario Schweitzer, Benny Trakhtenbrot and Ido Finkelman
for useful discussions. We also thank Benny Trakhtenbrot for retrieving the SDSS data used in this paper.
The referee, A. Lawrence, made useful comments that helped us to improve the paper.
Funding for this work has been provided by the Israel Science Foundation grant 364/07 and
a DIP grant 15981.
\end{acknowledgements}

\end{document}